\documentclass[twocolumn,showpacs,prb,aps,10pt]{revtex4-1}

\usepackage{graphicx}
\usepackage{dcolumn}
\usepackage{bm}

\newcommand{\I}{{\rm i}}

\begin{document}

\title{Excited states in poly-diacetylene chains:\\
A Density-matrix-renormalization-group study}

\author{Gergely Barcza$^1$}
\author{William Barford$^2$}
\author{Florian Gebhard$^3$}
\author{\"Ors Legeza$^{1,3}$}
\affiliation{$^1$ Strongly Correlated Systems {\sl Lend\"ulet\/}
Research Group, Wigner Research Centre,
Hungarian Academy of Sciences, H-1121 Budapest, Hungary\\
$^2$Department of Chemistry, University of Oxford, Oxford, OX1 3QZ, 
United Kingdom\\
$^3$Department of Physics and Material Sciences Center,
Philipps-Universit\"at D-35032 Marburg, Germany}

\date{March 11, 2013} 

\begin{abstract}
We study theoretically poly-diacetylene chains diluted in their monomer matrix.
We employ the density-matrix renormalization group method (DMRG)
on finite chains to calculate the ground state and low-lying excitations
of the corresponding Peierls--Hubbard-Ohno Hamiltonian
which is characterized by the electron transfer amplitude~$t_0$
between nearest neighbors, by the electron-phonon
coupling constant~$\alpha$, by the Hubbard interaction~$U$,
and by the long-range interaction~$V$.
We treat the lattice relaxation in the adiabatic limit,
i.e., we calculate the polaronic lattice distortions
for each excited state. Using chains with up to 102~lattice sites,
we can safely perform the extrapolation to the thermodynamic
limit for the ground-state energy and conformation, 
the single-particle gap, and the energies of
the singlet exciton, the triplet ground state, and
the optical excitation of the triplet ground state.
The corresponding gaps are known with high precision from experiment.
We determine a coherent parameter set
$(t_0^*=2.4\, {\rm eV}, \alpha^*=3.4\, {\rm eV}/\hbox{\AA}, 
U^*=6\, {\rm eV}, V^*=3\, {\rm eV})$ from a fit of the 
experimental gap energies to the theoretical values 
which we obtain for
81~parameter points in the four dimensional search space
$(t_0, \alpha, U, V)$.
We identify dark in-gap states in the singlet and triplet sectors
as seen in experiment.
Using a fairly stiff spring constant, 
the length of our unit cell is about one percent larger than
its experimental value.
\end{abstract}

\pacs{71.20.Rv, 71.10.Fd, 78.30.Jw, 78.20.Bh}

\maketitle

\section{Introduction}
\label{sec:intro}

Poly-diacetylene (PDA) chains dispersed with low concentration
in their monomer single-crystal matrix are prototypical quasi one-dimensional 
materials.~\cite{PDAreview,Sariciftci,Schott-review}
The structural disorder in the chains 
and their surrounding matrix is tiny,
the electronic excitation energies of the diacetylene monomers
are much higher than those of the polymer, and the
chains' electronic excitations in the energy range of visible light
can be measured with a very high accuracy.\cite{Schottzucht}

Exciton-polaritons have been generated that have been
shown to be coherent over tens of micrometers, i.e., several
ten thousand monomer units.~\cite{Schottexciton}
This observation was confirmed by weight measurements after dissolving the
chains and their monomer matrix.\cite{Schott-review} 
Consequently, the opto-electronic properties of the PDAs 
result from the electrons' mutual interaction
and their interaction with the lattice potential,
while the influence of disorder is negligible.
This makes these materials the perfect testing-ground for 
theoretical model studies which describe interacting electrons
on perfectly ordered chains.


The typical single-particle gap in PDAs is $E_{\rm gap}\gtrsim 2.4\, {\rm eV}$,
see Sect.~\ref{sec:exp}.
Density-functional theory band structure calculations 
in the local-density approximation for generic PDA geometries
estimate the bare band-gap to be 
$E_{\rm bare\, gap}\approx 1.2\, {\rm eV}$,
or less.~\cite{Parry}
Results from various methods are compiled in table~1 of
Ref.~[\onlinecite{YangandKertesz}];
recent calculations using the
Perdew-Burke-Ernzerhof global hybrid density functional 
and the 6-311G(2$d$,2$p$) basis set of
atom-centered Gaussian functions (geometries from the TPSS density functional
and the 6-31G($d$) basis set)~\cite{Janesko-paper}
give $E_{\rm bare\, gap}\approx 1.6\, {\rm eV}$.~\cite{Janesko-private} 
The comparison shows that 
electronic exchange and correlations account for a substantial fraction
of the single-particle gap. 
In contrast to inorganic semiconductors,
the exciton binding energy in PDAs amounts to about
20\% of the single-particle gap. Such large binding energies suggest that
the electron-electron interaction must be treated accurately for the
calculation of the optical properties of the PDAs.

In order to describe the optical excitations in PDAs,
two approaches have been taken. The first approach starts
from an ab-initio density-functional theory calculation
of the bare band structure in local-density approximation (LDA),
which is then supplemented by an approximate treatment of the residual
electron-electron interaction, e.g., the $GW$ approximation 
for the single-particle bands and the Bethe-Salpeter equation (BSE)
for the excitons (LDA+$GW$+BSE).~\cite{LouieRohlfing,LDA-GW-BSE}
Actual calculations for the PDAs often omit the $GW$~step 
(Wannier theory).~\cite{van-der-Horst}
Within this approach, a number of experimental data can be 
reproduced, e.g., the ${1}^1B_u$ exciton binding energy
and its polarizability.

This approach is less successful for the triplet sector.~\cite{van-der-Horst}
Typically, the energy of the triplet ground state is too high.
Recall that, if the electron-electron interaction is absent,
the energy of the triplet ground state is identical to
the single-particle gap. Starting from a weak-coupling
description of the electron-electron interaction on the chains, 
it is difficult to obtain the experimentally 
observed energy renormalization by a factor of almost 
three.~\cite{Rissler-Grage}
Moreover, polaronic effects are not considered in the LDA+BSE approach.

The second approach to a theoretical description
of the primary excitations in polymers starts from
a many-particle model Hamiltonian that describes
only the $\pi$-electrons and their mutual interaction.
Typically, empirical parameters are used
for the tight-binding band structure
and for the Pariser-Parr--Pople (PPP) potential.~\cite{PPP}
With the help of the 
density-matrix renormalization-group (DMRG)
method,\cite{steve} the ground state and elementary excitations
for such models can be calculated for large chains with very high
accuracy. In this way, the electron-electron interaction
is treated without resorting to any approximations.

In a recent study,\cite{all-of-us} we used
the Hubbard-Ohno potential and the tight-binding parameters
of Ref.~[\onlinecite{Bursill2001}] to calculate
the binding energy, polarizability and wave function of
the singlet exciton, in good agreement with experiment.
However, in our previous study we 
could not reproduce satisfactorily the energy of the
triplet ground state. Moreover, we did not
take polaronic effects into account.
Here, we shall overcome these shortcomings.

In this work, we perform an extensive DMRG study of 
the Peierls--Hubbard-Ohno Hamiltonian for the $\pi$-electrons
on a chain. We start from the tight-binding Peierls description of
Race et al.~\cite{Bursill2003} but replace their Ohno potential~\cite{Ohno}
by the Hubbard-Ohno potential.\cite{all-of-us}
The essential difference between the two parameterizations of the
Pariser-Parr--Pople interaction~\cite{PPP} lies in the treatment of the
Coulomb interaction for $\pi$-electrons on the carbon atom.
The local Hubbard repulsion potential is substantially larger than 
the corresponding Ohno interaction.~\cite{Chandross}

Since we are mostly interested in the polaronic effects,
we ignore the energetic effects introduced by the ligands~$R,R'$.
Some preliminary studies show that the energy difference
between the singlet and triplet ground states is not influenced
by the introduction of a local potential for the carbon atoms
which are linked to the side groups.
We presume that the dominant influence of the side groups 
comes from the presence (or absence) of strain in the chains.
As can be seen from the data for 3BCMU and 4BCMU,
different ligands and the resulting strain 
results in small energetic differences, of about $0.1\, {\rm eV}$
for the nBCMU family.

The outline of our work is as follows.
In Sect.~\ref{sec:exp}, we summarize the experimental observations
on the singlet and triplet in-gap states in PDAs. 
In Sect.~\ref{sec:model}, we define the Peierls--Hubbard-Ohno
Hamiltonian and the model parameters
which provide the basis of our numerical DMRG study.
In Sect.~\ref{sec:method}, we briefly discuss our numerical approach.
In Sect.~\ref{sec:guessparameters}, we motivate the parameter regime
that we choose for our study.
In Sect.~\ref{sec:results}, we present our results.
In Sect.~\ref{sec:conclusions}, we summarize and conclude.
Technical details are deferred to the appendices.

\section{Experimental observations}
\label{sec:exp}

We start with an overview of the experimental observations
relevant for our study.

\subsection{Ground-state conformation}
\label{subsec:structure}

First, we collect relevant experimental data on the ground-state
properties.

\subsubsection{Lewis structure}

The diacetylene monomer building unit is comprised of four carbon atoms.
The four outer electrons of each carbon atom are hybridized.
Three of them form localized bonds. There are $\sigma$-bonds between
neighboring carbon atoms on the chain.
Two carbon atoms on the chain share a local $\pi$-bond made by two 
$p_y$~electrons. The other two carbon atoms share $\sigma$-bonds
to covalent ligands 
$R$ and $R'$, which are several \AA ngstr\o m long and differ for
various members of the PDA family.
In this work, we focus on poly-(Butoxy-Carbonyl-nMethylene-Urethane)
(poly-nBCMU) chains with $n=3,4$ where the side groups are given by
$R = R' = 
({\rm CH}_2)_n-{\rm OCONH}-{\rm CH}_2-{\rm COO}-({\rm CH}_2)_3{\rm CH}_3$.

\begin{figure}[htb]
\includegraphics[height=2.4cm]{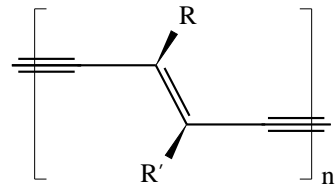}
\caption{Lewis structure of a poly-diacetylene unit cell.\label{fig:structure}}
\end{figure}

The fourth carbon electron is delocalized over the carbon backbone 
in a molecular $\pi$-orbital. 
Due to the Peierls effect, in the ground state the $\pi$~electrons
dimerize the chain into an alternating sequence of short and long bonds.
After dimerization, the four carbon atoms in 
the unit cell are linked by a triple bond, 
a single bond, a double bond, and a single bond.
The corresponding Lewis structure of the ground state
is shown in Fig.~\ref{fig:structure}.

\subsubsection{Lattice parameters}
\label{subsubsec:latticeparameters}

For a high-quality single crystal of
Poly-[1,2-bis-(p-tolylsulfonyloxymethylen)-1-buten-3-inylen] (PTS),
the atomic distances at room temperature have been measured~\cite{KobeltPaulus}
as $r_{\rm t}=1.191(4)\, \mbox{\AA}$, 
$r_{\rm d}=1.356(4)\, \mbox{\AA}$, 
and $r_{\rm s}=1.428(4)\, \mbox{\AA}$
[uncertainties in the last digit in brackets]
for the triple~(t), double~(d), and single~(s) bonds, 
respectively.
Typical atomic distances for other PDA polymer single crystals 
are~\cite{ChemlaZyss} $r_{\rm t}=1.20\, \mbox{\AA}$,
$r_{\rm d}=1.36\, \mbox{\AA}$, and 
$r_{\rm s}=1.43\, \mbox{\AA}$.
The same set of data applies for
3BCMU-PDA at low temperatures.\cite{SpagnoliRaman} 
The chain of atoms is not straight; the single and double bonds
alternately form angles of $\varphi_1=120^{\circ}$ and $\varphi_2=240^{\circ}$ 
degrees, with a temperature variation of a few degrees.~\cite{SpagnoliRaman}
In the comparison with our calculations we shall assume 
that the bond lengths and angles given above for the PDA single crystals
are representative for chains in their diacetylene monomer matrix.

We shall only deal with planar (`blue') 
PDA chains.~\cite{Schott-review} 
The individual polymer chains of 4BCMU are
strained in their monomer single crystals but are
essentially unstrained in 3BCMU.~\cite{SpagnoliRaman}
Strain should be the primary source for differences
in the spectra of these two PDAs.

\subsection{Excited states}

Next, we summarize experimental results on the low-lying
electronic excitations.
PDAs are center-symmetric insulators. Their ground state~G
is a spin singlet with symmetry $A_g$
under inversion. 

\subsubsection{Single-particle gap and excited singlet states}
\label{subsec:opticalprops}

The charge gap for single-particle excitations,
as determined from Franz--Keldysh oscillations in electro-absorption
experiments,\cite{HorvathWeiserLapersonne} is
$E_{\rm gap}({\rm 3BCMU})=2.482\, {\rm eV}$
and 
$E_{\rm gap}({\rm 4BCMU})=2.378\, {\rm eV}$
in 3BCMU and 4BCMU chains, respectively.

\begin{figure}[htb]
\vspace*{9pt}
\hspace*{2pt}\includegraphics[width=0.95\columnwidth]{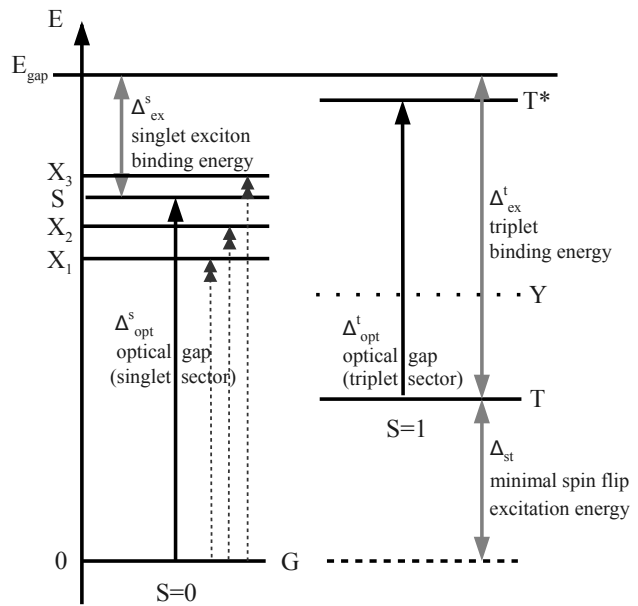}
\caption{Energy levels of in-gap states in the spin-singlet 
and spin-triplet sectors.
Single-tip arrows: optical
absorption spectroscopy; 
double-tip arrows: two-photon absorption spectroscopy.
Double arrows: binding energies (gaps).
G:~singlet ground state ($1{}^1A_g$); S: singlet exciton ($1{}^1B_u$);
X$_1$, X$_2$, X$_3$: singlet dark states ($m{}^1A_g$);
T: triplet ground state ($1{}^3B_u$); T$^*$: optical
excitation of the triplet ground state ($1{}^3A_g$);
Y: dark triplet state ($m{}^3B_u$).\label{fig:energylevels}}
\end{figure}

The excitation energy of the primary singlet exciton~S 
(symmetry ${}^1B_u$) defines the optical gap, 
$\Delta_{\rm opt}^{\rm s}=E_{\rm S}-E_{\rm G}$, which amounts to
$\Delta_{\rm opt}^{\rm s}({\rm 3BCMU})= 1.896\, {\rm eV}$
and $\Delta_{\rm opt}^{\rm s}({\rm 4BCMU})= 1.810\, {\rm eV}$
in 3BCMU and 4BCMU, respectively.
Therefore, the singlet exciton binding energy, 
defined by $\Delta_{\rm ex}^{\rm s}= 
E_{\rm gap}-E_{\rm S}$, becomes 
$\Delta_{\rm ex}^{\rm s}({\rm 3BCMU})= 0.586\, {\rm eV}$ in 3BCMU,
and $\Delta_{\rm ex}^{\rm s}({\rm 4BCMU})= 0.568\, {\rm eV}$ in 4BCMU,
about 24\% of the band-gap. 
The energy levels are sketched in Fig.~\ref{fig:energylevels}.

The singlet exciton is the energetically lowest state in the
spin-singlet sector which
can be generated by a single-photon absorption process.
In addition, there are further, optically dark states in the gap.
The existence of optically dark states~X$_1$ and X$_2$
below the optical gap can
be inferred from non-radiative decay processes which are monitored
via pump-probe spectroscopy, see Ref.~[\onlinecite{Schott-review}]
for a review. The exciton rapidly populates the states X$_{1,2}$
so that they should have the same spin quantum number.

In principle, the energy of singlet states in the gap with
${}^1A_g$-symmetry can be determined via two-photon absorption.
Two-photon absorption for a single-crystal of 
the poly-diacetylene paratoluene-sulfonate reveals
three gap states with energies $E_{\rm X_3}=1.05 \Delta_{\rm opt}^{\rm s}$,
$E_{\rm X_2}=0.9 \Delta_{\rm opt}^{\rm s}$, and
$E_{\rm X_1}=0.8 \Delta_{\rm opt}^{\rm s}$. 
These states should exist in all PDAs.
Note that in the experimental literature,~\cite{Schott-review}
the numbering of the states X$_1$ and X$_2$ is reversed.

\subsubsection{Phonon energies}
\label{subsec:phononreplicas}

Raman scattering reveals vibrational energies which are assigned
to the oscillations of the double (D) and triple (T) bonds.
For 3BCMU chains in their monomer matrix they are
$\hbar \omega_{\rm D}=0.181\, {\rm eV}$ 
and $\hbar \omega_{\rm T}=0.261\, {\rm eV}$.~\cite{SpagnoliRaman}
In accordance with the Raman data, the optical spectra of 3BCMU 
chains show strong exciton replicas at the energies 
$E_c=E_{\rm S}+\hbar\omega_{\rm D}=2.079\, {\rm eV}$
and at 
$E_d=E_{\rm S}+\hbar\omega_{\rm T}=2.160\, {\rm eV}$, respectively,
when the exciton is accompanied by single optical phonons
corresponding to the vibrations with frequencies
$\omega_{\rm D}$ and $\omega_{\rm T}$.

Electro-absorption measurements~\cite{Schott-review,Weiserprivate} show
that there are more significant
single-phonon replicas of the singlet exciton 
at the energies $E_{a,b}=E_{\rm S}+\hbar\omega_{\rm S,D^*}$ 
with $\hbar\omega_{\rm S}=0.090\, {\rm eV}$ and
$\hbar\omega_{\rm D^*}=0.155\, {\rm eV}$.
Due to the fact that $E_b\approx E_c$,
the phonon replica at $E_b$ appears in the
low-energy flank of the replica at energy $E_c$.
The electro-absorption measurements also permit the
identification of multi-phonon replicas, e.g., 
at replica energies $\hbar(\omega_{\rm D^*}+\omega_{\rm D})$,
$2\hbar\omega_{\rm D}$, $\hbar(\omega_{\rm D}+\omega_{\rm T})$,
$2\hbar\omega_{\rm T}$.~\cite{Weiserprivate}

\subsubsection{Triplet ground state and excited triplet states}
\label{subsec:opticalpropstriplet}

The triplet sector is more difficult to access experimentally
because a transition between the spin-zero ground state~G and
the lowest spin-one state~T is optically forbidden.
Optical pump-probe spectroscopy~\cite{Schott-review}
reveals that a small fraction of singlet excitons decays 
into a long-lived state which can be optically excited
by the probe pulse.
Its long life-time indicates that this in-gap state 
is the spin-triplet ground state~T.
The probe pulse generates transitions
from~T to T$^*$ in the triplet sector.
The optical gap in the triplet sector is defined as
$\Delta_{\rm opt}^{\rm t}=E_{\rm T^*}-E_{\rm T}$, and amounts to
$\Delta_{\rm opt}^{\rm t}({\rm 3BCMU})=1.360\, {\rm eV}$
and 
$\Delta_{\rm opt}^{\rm t}({\rm 4BCMU})=1.345\, {\rm eV}$,
respectively.

Optical pumping above a threshold $E_{\rm f}\approx 2.0\, {\rm eV}$ 
very efficiently generates states 
which show a strong optical absorption with energy~$\Delta_{\rm opt}^{\rm t}$.
This can be readily understood if the spin-singlet
excitations above $E_{\rm f}$ fission into 
triplet pairs.~\cite{Schott-review} 
In turn, these triplet pairs can recombine 
into singlets and decay optically.
If we ignore lattice effects (bi-polaron formation),
we obtain a reasonable estimate for the energy
of the triplet state~T,
$\Delta_{\rm st}=E_{\rm T}-E_{\rm G}\approx E_{\rm f}/2$.
For 3BCMU and 4BCMU this estimate gives
$\Delta_{\rm st}({\rm 3BCMU})\gtrsim 1.0\pm 0.05\, {\rm eV}$
and $\Delta_{\rm st}({\rm 4BCMU})\gtrsim 0.95\pm 0.05\, {\rm eV}$, 
respectively.

The binding energy of the triplet ground state is defined by
$\Delta_{\rm ex}^{\rm t}=E_{\rm gap}-E_{\rm T}$. It amounts to
$\Delta_{\rm ex}^{\rm t}({\rm 3BCMU})\approx 1.5\, {\rm eV}$
and $\Delta_{\rm ex}^{\rm t}({\rm 4BCMU})\approx 1.4\, {\rm eV}$ 
in 3BCMU and 4BCMU, 
respectively, more than 60\% of the single-particle gap. 
The energy of the optically excited triplet ground state~T$^*$ 
is found to be
$E_{{\rm T}^*}({\rm 3BCMU})-E_{\rm G}=2.36\,  {\rm eV}$ and
$E_{{\rm T}^*}({\rm 4BCMU})-E_{\rm G}=2.30\,  {\rm eV}$
above the ground state, about $0.1\, {\rm eV}$
below the threshold $E_{\rm gap}$ for single-particle excitations.

The optically dark singlet states X$_1$ and X$_2$ decay non-radiatively
into a state~Y. Its weak population and long life-time
indicate that it is reached via intersystem crossing
so that it ought to be a spin-triplet state which lies energetically
above the triplet ground state~T. It should be of
symmetry ${}^3B_u$ so that it cannot be reached via an
optical excitation of the triplet ground state.

Fig.~\ref{fig:energylevels} shows the experimentally observed level
spectrum for PDA chains in the singlet and triplet sectors.
We summarize the corresponding values for the in-gap states
in table~\ref{tab:BCMU-energies}, and compare them to our theoretical results
for our best parameter set,
$(t_0^*=2.4\, {\rm eV}, \alpha^*=3.4\, {\rm eV}/\hbox{\AA}, 
U^*=6\, {\rm eV}, V^*=3\, {\rm eV})$, see Sect.~\ref{sec:results}.

\begin{table}[htb]
\begin{center}
{\tabcolsep=3pt\begin{tabular}{|l|l|l|l|}
\colrule\vphantom{\LARGE A}Energy & 3BCMU & 4BCMU &
Theory\\
\colrule\vphantom{\LARGE A}%
$E_{\rm X_1}$ &  {\sl 1.5} &  {\sl 1.4} & 1.74 [1.94]\\
$E_{\rm X_2}$ &  {\sl 1.7} &  {\sl 1.6} & 1.85 [1.94]$^{\rm a}$\\
$E_{\rm S}=\Delta_{\rm opt}^{\rm s}$ &  {\bf 1.896} &  {\bf 1.810} & 2.00 [2.05]\\
$E_{\rm X_3}$ &  {\sl 2.0} &  {\sl 1.9} & \\
$E_{\rm gap}$ &  {\bf 2.482} & {\bf 2.378} & 2.45 [2.47]\\
$\Delta_{\rm ex}^{\rm s}=E_{\rm gap}-E_{\rm S}$
&{\bf 0.586} &{\bf 0.568} & 0.45 [0.42]\\[1pt]
\colrule\vphantom{\LARGE A}%
$E_{\rm T}=\Delta_{\rm st}$ 
&  {\sl 1.0\hphantom{00}} $\pm$ {\sl 0.05} 
&  {\sl 0.95\hphantom{0}} $\pm$ {\sl 0.05} 
& 1.00 [1.06]\\
$E_{{\rm T}^*}=\Delta_{\rm st}+\Delta_{\rm opt}^{\rm t} $
& {\sl 2.36\hphantom{0}} $\pm$ {\sl 0.05} 
& {\sl 2.30\hphantom{0}} $\pm$ {\sl 0.05} 
& 2.25\\
$\Delta_{\rm opt}^{\rm t}$ &  {\bf 1.360} &  {\bf 1.345}& 1.25 [1.28]\\
$\Delta_{\rm ex}^{\rm t} =E_{\rm gap}-E_{\rm T}$
&  {\sl 1.5\hphantom{00}} $\pm$ {\sl 0.05} 
&  {\sl 1.4\hphantom{00}} $\pm$ {\sl 0.05}
&  1.45 [1.40] \\[1pt]
\colrule
\end{tabular}
\vspace*{-6pt}
\footnotetext{Several degenerate states are found in DMRG.}
}
\end{center}
\caption{First and second column:
Excitation energies in 3BCMU and 4BCMU at low temperatures.
All energies are measured in eV relative to the energy
of the ground state, $E_{\rm G}=0$.
Bold number: directly measured; italic number: estimate.
Third column: our results, see Sect.~\ref{sec:results};
the numbers in square brackets give the excitation energy
for the rigid-lattice transition from~G 
($E_{\rm gap}$, $E_{\rm S}$, $E_{{\rm X}_{1,2}}$, $E_{\rm T}$)
and from~T ($\Delta_{\rm opt}^{\rm t}$).\label{tab:BCMU-energies}}
\end{table}

\section{Model description of poly-diacetylene chains}
\label{sec:model}

In this work we  
restrict ourselves to the description of the $\pi$~electrons
because they dominate the optical response of the poly-diacetylene
chains immersed in their monomer matrix 
for energies $\hbar \omega<3\, {\rm eV}$.
In order to make contact with previous work,~\cite{Bursill2003}
we treat the other electrons as inert, i.e.,
they are supposed to form the {\em unrelaxed\/} geometry of the carbon backbone.
The distance between two carbon atoms is $R_2$ for a single
$\sigma$-bond, and $R_1<R_2$ for the $\sigma$-$p_y$ double bond
(extrinsic distortion).

\subsection{Electronic Hamiltonian}

The motion of $\pi$-electrons between neighboring carbon atoms
and their mutual Coulomb interactions defines the electronic problem,
\begin{equation}
\hat{H}_{\rm e}=\hat{T}+\hat{V} \; ,
\end{equation}
where $\hat{T}$ and $\hat{V}$ specify the electrons' kinetic energy and
their mutual interaction, respectively.

\subsubsection{Kinetic energy}
\label{subsec:kinetic}

The motion of the $\pi$-electrons over the unrelaxed backbone
is described by the operator for the kinetic energy,
\begin{equation}
\hat{T} = - \sum_{l;\sigma} t_{l} 
\left( \hat{c}_{l,\sigma}^+\hat{c}_{l+1,\sigma}
+ \hat{c}_{l+1,\sigma}^+\hat{c}_{l,\sigma}\right) \; ,
\label{eqn:hatT}
\end{equation}
where $\hat{c}^+_{l,\sigma}$,
$\hat{c}_{l,\sigma}$ are creation and annihilation operators, respectively, for
a $\pi$-electron with spin~$\sigma=\uparrow,\downarrow$ on site~$l$
with two-dimensional coordinate $\vec{r}_{l}=(x_l,y_l)^T$.
The matrix elements $t_{l}$ are the electron transfer 
amplitudes between neighboring
sites. Transfer amplitudes between next-nearest neighbors
should be included to fit better the band structure of all carbon 
electrons.~\cite{Janesko-private}
The amplitude for an electron transfer 
between two carbon sites at distance $r_0=1.4\, \mbox{\AA}$
is given by $t_0$ which we use as an adjustable parameter.

We consider the half-filled band exclusively, i.e.,
in the ground state and for the excitations in Fig.~\ref{fig:energylevels}
the number of $\pi$~electrons~$N_{\rm e}$ equals the
number of lattice sites~$N$. 

\subsubsection{Coulomb interaction}
\label{subsec:Coulomb}

The diacetylene monomer single-crystals are insulators,
and also the PDA chains display a finite charge gap. 
Therefore, the long-range Coulomb
interaction is not dynamically screened at the energy scale
of a few electron volts.

Therefore, we start from the Pariser-Parr--Pople (PPP) interaction~\cite{PPP} 
\begin{eqnarray}
\hat{V} &=& 
U \sum_{l=1}^N \left(\hat{n}_{l,\uparrow}-\frac{1}{2}\right)
\left(\hat{n}_{l,\downarrow}-\frac{1}{2}\right) \nonumber \\
&& + \frac{1}{2}\sum_{l\neq m=1}^N V_{l,m}^{\rm PPP}
\left[\left(\hat{n}_{l}-1\right)
\left(\hat{n}_{m}-1\right)\right]\; .
\label{eqn:ourinteraction}
\end{eqnarray}
Here, $\hat{n}_{l}=\hat{n}_{l,\uparrow}
+\hat{n}_{l,\downarrow}$ counts the number of electrons on site~$l$,
and $\hat{n}_{l,\sigma}= \hat{c}^+_{l,\sigma}\hat{c}_{l,\sigma}$ 
is the local density operator at site~$l$ for 
spin~$\sigma$. 
The strength of the (local) Hubbard interaction is parameterized
by~$U$, and $V_{l,m}$ are the PPP parameters for the effective Coulomb repulsion
between electrons at different positions $\vec{r}_l$ and $\vec{r}_m$.

For the description of electrons and holes in quantum wires
and other quasi one-dimensional structures in vacuum,
various effective potentials 
have been used in the literature.~\cite{Loudon,KochundCo}
For example, in our previous study~\cite{all-of-us} we used
the erf-potential ($x=|\vec{r}_l-\vec{r}_m|$)
\begin{equation}
V_{l,m}^{\rm erf}=V^{\rm erf}(x)=\frac{e^2}{\epsilon_d R}\sqrt{\pi} \exp[(x/R)^2]
\left[1-{\rm erf}(x/R)\right]\; ,
\end{equation}
where ${\rm erf}(x)$ is the error function,
$R$ is the adjustable confinement parameter, and $\epsilon_d=2.3$ 
is the static dielectric constant for the diacetylene monomer matrix.
In general, the PPP interaction for the $\pi$-electrons on the chain
in the surrounding matrix has the form
\begin{equation}
V_{l,m}^{\rm PPP}=V^{\rm PPP}(x)=V^{\rm erf}(x) \frac{\epsilon_d}{\epsilon(x)} \; ,
\end{equation}
where $\epsilon(x)$ is the static dielectric function at distance
$x=|\vec{r}_l-\vec{r}_m|$ with $\epsilon(x\to\infty)=\epsilon_d$.
Unfortunately, the short-distance behavior of $\epsilon(x)$
is unknown. 

In this work 
we follow Refs.~[\onlinecite{all-of-us,Bursill2001,Chandross}] and approximate
the Pariser-Parr--Pople interaction
using the Hubbard-Ohno potential, i.e., for $x=|\vec{r}_l-\vec{r}_m|\neq 0$
we set
\begin{eqnarray}
V_{l,m}^{\rm PPP} \approx V^{\rm Ohno}_{l,m}&=& V^{\rm Ohno}(x)= 
\frac{V}{\epsilon_d\sqrt{1+\beta(x/\mbox{\AA})^2}} \; ,\nonumber \\ 
\beta &=& \left(\frac{V}{14.397\, {\rm eV}}\right)^2 \; .
\label{eq:Ohno-potential}
\end{eqnarray}
The Ohno potential and its adjustable parameter~$V$ 
describe the effective strength of the Coulomb interaction 
at short distances; for large electron-electron distances, 
$V^{\rm Ohno}(x\to\infty)\to e^2/(\epsilon_d x)$
because $e^2= 14.397\, {\rm eV}\, \mbox{\AA}$.
Note that $U$ and $V$ are independent adjustable parameters
in our theory. Later, we shall assume that the screening of the 
on-site interaction is substantially less effective 
than for the long-range interaction.

The eigenstates of the electronic problem
follow from the solution of the corresponding
Schr\"odinger equation,
\begin{equation}
\hat{H}_{\rm e} | \Psi \rangle = E_{\Psi} | \Psi \rangle \; .
\end{equation}
We denote expectation values of operators $\hat{A}$
in the normalized state $|\Psi\rangle$
as $\langle \hat{A} \rangle=\langle \Psi | \hat{A} | \Psi \rangle$.

\subsubsection{Particle-hole symmetry}
\label{subsec:phtransform}

The Hamiltonian~(\ref{eqn:hamilt}) is
invariant under the particle-hole transformation 
$\hat{c}_{l,\sigma}\mapsto (-1)^l\hat{c}_{l,-\sigma}^+$. 
At half band-filling, 
the ground state $|{\rm G}\rangle$
is also invariant under this transformation. 

In our numerical investigation, we calculate important 
excited states using the proper quantum numbers for the spin 
symmetry ($S=0,1$; $p=2S+1$),
inversion symmetry ($X=A_g, B_u$), and particle-hole symmetry
($v=\pm 1$).
Therefore, we label states in the form $m\; {}^p X^v$ 
where $m\ge 1$ counts the states 
with the same symmetry in ascending energetic order. 
The quantum numbers for the most prominent
in-gap states are summarized in table~\ref{tab:qnumbers}.

\begin{table}[htb]
\begin{tabular}{|l|c|c|c|c|}
\colrule
&\multicolumn{4}{c|}{\vphantom{\LARGE A}State}\\
\colrule
\vphantom{\LARGE A}Symmetry &  G &  T&  S&  T$^*$\\
\colrule\vphantom{\LARGE A}Spin ($S$) 
& \hphantom{$-$}0 & \hphantom{$-$}1 & \hphantom{$-$}0 & \hphantom{$-$}1 \\
Inversion ($X$) 
& \hphantom{$-$}1 & $-$1 & $-$1 & \hphantom{$-$}1 \\
Particle-hole ($v$) 
& \hphantom{$-$}1 & \hphantom{$-$}1 & $-$1 & $-$1 \\
\colrule
\vphantom{\LARGE A}Classification
& $1\; {}^1 A_g^+$  & $1\; {}^3 B_u^+$  & $1\; {}^1 B_u^-$ & $1\; {}^3 A_g^-$ \\
\colrule
\end{tabular}
\caption{Quantum numbers of important in-gap states.
For the definition of the states, see Fig.~\protect\ref{fig:energylevels}.
\label{tab:qnumbers}}
\end{table}
   
\subsection{Electron-lattice interaction}

The electron-phonon coupling leads to the dimerization of the
ground-state structure (Peierls effect). Moreover,
excitations carry a polaron cloud, and the
polaronic shifts in the single-particle excitation energies
will in general be different from those of bound pairs.
For example, the singlet exciton fissions
into a bound pair of triplet polarons. This bipolaron has an energy
which is lower than twice the energy of a triplet polaron.
Therefore, the estimates in table~\ref{tab:BCMU-energies}
provide lower bounds for $\Delta_{\rm st}$.

Due to the (small) Peierls distortion, the energy increases 
so that the total Hamiltonian reads
\begin{equation}
\hat{H}=\hat{H}_{\rm e} + \sum_{l=1}^{N-1}\frac{\delta_l^2}{4\pi t_0\lambda_l}
\; .
\label{eqn:hamilt}
\end{equation}
At distance $r_0=1.4\, \mbox{\AA}$, the electron transfer matrix element 
is given by our adjustable parameter $t_0$.
The strength of the electron-lattice coupling is
parameterized by the coupling constant~\cite{SSH,Ehrenfreund}
\begin{equation}
\lambda_l = \frac{2\alpha^2}{\pi K_l t_0}
\end{equation}
where $\alpha$
is the strength of the Peierls coupling and
$K_l$ are the elastic constants for the carbon backbone,
namely, $K$~for the $\sigma$-bonds and $G>K$ for the $\sigma$-$p_y$ bonds,
$K_l=K+\delta_{l\, {\rm mod}\, 4,1}(G-K)$.
The parameters $\alpha$, $K$, and $G$ must be adjusted.
In effect, we address the region where $\lambda_l<0.1$ 
so that the adiabatic approximation is valid, i.e.,
we may treat the lattice distortions classically.

The intrinsic Peierls distortion  
implies a modulation of the bond lengths,
\begin{equation}
\delta r_l=r_l- R_l= -\frac{\delta_l}{2\alpha} \; .
\label{eq:bondlengths}
\end{equation}
Here, $R_2=r_0$ for a single $\sigma$-bond of the carbon backbone chain and
$R_1=r_0-\delta^{\rm e}/(2\alpha)$ for its $\sigma$-$p_y$ double bond.
The size of the extrinsic dimerization $\delta^{\rm e}$
is calculated in App.~\ref{appA}.
As a result of the intrinsic Peierls dimerization, 
the ground-state unit cell of the distorted chain consists
of four carbon atoms, linked by a single bond, a double bond, a single 
bond, and a triple bond, see Fig.~\ref{fig:structure}.

The energies $\delta_l$ modulate the electron transfer amplitudes,
\begin{equation}
t_l= \overline{t}_l + \delta_l/2
\end{equation}
with 
\begin{eqnarray}
\overline{t}_l &=&t_0+\delta^{\rm e}/2 \nonumber \\ 
&& \hbox{(undistorted $\sigma$-$p_y$ double bond),} \\
\overline{t}_l &=&t_0\nonumber \\
&& \hbox{(undistorted $\sigma$ single bond).}
\end{eqnarray}
Note that, for small distortions,
the transfer matrix elements
for the single, double, and triple bonds in the ground state obey
\begin{equation}
t_{\rm s,d,t }=t_0 -\alpha(r_{\rm s,d,t}-r_0)\; , 
\label{eq:Tparameters}
\end{equation}
where $r_{\rm s}$, $r_{\rm d}$, and $r_{\rm t}$ are the lengths of the
single, double, and triple bond in the unit cell, 
see eq.~(\ref{eq:bondlengths}).

\subsection{Model parameters}

Our model employs the following parameters:
(i)~The electron transfer matrix element $t_0$ for a single C--C-bond 
at distance $r_0=1.4\, \mbox{\AA}$;
(ii)~the strength of the local Hubbard interaction~$U$;
(iii)~the strength of the short-range Coulomb interaction~$V$;
(iv)~the Peierls coupling~$\alpha$;
(v)~the spring constants~$K$ and~$G$.
The model parameters are adjusted to reproduce 
the single-particle gap~$E_{\rm gap}$, the singlet exciton energy $E_{\rm S}$,
the energy of the triplet state~$E_{\rm T}$,
and the optical gap in the triplet sector~$\Delta_{\rm opt}^{\rm t}$,
see table~\ref{tab:BCMU-energies}.
Moreover, we estimate the values for the spring constants
from the energy of the phonon replicas.

Of course, we cannot scan a five-dimensional parameter space
completely. Therefore, we have to restrict ourselves to
values which seem plausible, see Sect.~\ref{sec:guessparameters}.
The model parameters investigated
are summarized in table~\ref{tab:parameters}.

\begin{table}[ht]
\begin{tabular}{@{}ll@{}}
\begin{tabular}[t]{|c|c|}
\colrule\begin{tabular}{@{}c@{}}\vphantom{\LARGE A}Fixed\\ parameter 
\end{tabular}& Value\\
\colrule\vphantom{\LARGE A}$\epsilon_d$ & 2.3 \\
$K$ & 44 eV/\AA$^2$ \\
$G$ & 68 eV/\AA$^2$ \\
$r_0$ & 1.4 \AA\\[3pt]
\colrule
\end{tabular} &
\begin{tabular}[t]{|c|c|}
\colrule\begin{tabular}{@{}c@{}}\vphantom{\LARGE A}Control\\ parameter 
\end{tabular}& 
\begin{tabular}{@{}c@{}}\vphantom{\LARGE A}Range of\\ values
\end{tabular}\\
\colrule\vphantom{\LARGE A}$t_0$ & 2.0 eV \ldots 2.4 eV \\
$V$ & 2 eV \ldots 3 eV \vphantom{eV/\AA$^2$}\\
$U$ & 5 \ldots 6 eV \vphantom{eV/\AA$^2$}\\
$\alpha$ & 3.4 eV/\AA\ \ldots 3.6 eV/\AA\\[3pt]
\colrule
\end{tabular}
\end{tabular}
\caption{Parameters of the Peierls--Hubbard-Ohno model.\label{tab:parameters}}
\end{table}

\section{Method}
\label{sec:method}

First, we outline our procedure to find the optimal
lattice structure. Next, we define the single-particle gap
and describe how we address excited in-gap states.
Lastly, we remark on our DMRG procedure and the extrapolation
of our finite-size data to the thermodynamic limit.

In this section and in the remainder of the paper,
all energies ($t_0,U,V;E_{\rm gap},\Delta_{\rm opt}^{\rm s},\Delta_{\rm st},
\Delta_{\rm opt}^{\rm t}$) are given in eV, 
all lengths are given in \AA, 
and $\alpha$ is given in units of eV/\AA.

\subsection{Optimization of the lattice structure}

\subsubsection{Procedure}

The values for the electron transfer amplitude modulations follow
from minimization of the energy functional $E_{\Psi}(\delta_l)
=\langle \Psi |\hat{H} | \Psi \rangle$
for the normalized state $|\Psi\rangle$ which can be the ground state
or any excited state of $\hat{H}$.
The actual values for the dimerization are obtained from
the minimization of the energy functional $E_{\Psi}(\delta_l)$ with respect
to $\delta_l$ subject to the constraint 
\begin{equation}
\sum_{l=1}^{N-1}\delta_l=0 \; .
\label{eq:Gamma-condition}
\end{equation}
This reflects the fact that the total chain length should be fixed; see below.
The condition~(\ref{eq:Gamma-condition}) 
is taken into account with the help of the 
Lagrange multiplier~$\Gamma$. 

According to the 
Hellmann--Feynman theorem,~\cite{Solyom3} the negative derivatives
of the energy functional with respect to $\delta r_l$
define the force fields $f_l$,
\begin{eqnarray}
-\frac{f_l}{2\alpha} &=&  -\frac{\delta_l}{2\pi t_0\lambda_l} -\Gamma 
+ F_l[\delta_p]
\; ,\label{eq:def-f}\\
F_l[\delta_p]&=&  \frac{1}{2} \Bigl\langle 
\sum_{\sigma} \left( \hat{c}_{l,\sigma}^+\hat{c}_{l+1,\sigma}
+ \hat{c}_{l+1,\sigma}^+\hat{c}_{l,\sigma}\right) \Bigr\rangle \nonumber \\
&& + \frac{1}{2\epsilon_d}\sum_{i\neq j=1}^N 
\frac{\beta V/\mbox{\AA}{}^2}{[1+\beta 
(\left|\vec{r}_i-\vec{r}_j\right|/\mbox{\AA})^2\,]^{3/2}} 
\label{eq:forcefields}\\
&&
\times \left[x_{i,j} \frac{\partial x_{i,j}}{\partial \delta_l} 
+y_{i,j} \frac{\partial y_{i,j}}{\partial \delta_l} \right]
\langle \left(\hat{n}_{i}-1\right)
\left(\hat{n}_{j}-1\right)\rangle\; ,
\nonumber
\end{eqnarray}
where $\delta_p$ ($p=1,\ldots,N-1$)
are the Peierls modulations of the electron transfer amplitudes
and $x_{i,j}=x_i-x_j$, $y_{i,j}=y_i-y_j$.

For fixed bond angles, $x_l$ and $y_l$ are defined as
$x_l=x_{l-1}+r_l\cos(60^{\circ})$ and $y_l=y_{l-1}-r_l\sin(60^{\circ})$ 
for the bonds at an angle of $120^{\circ}$,
and $x_l=x_{l-1}+r_l$ and $y_l=y_{l-1}$ otherwise, 
with $r_l=R_l-\delta_l/(2\alpha)$ from~(\ref{eq:bondlengths}).
The force fields $f_l$ are zero at the optimal values $\delta_l^{\rm opt}$ for
a chosen state $|\Psi\rangle$. Note that $|\Psi\rangle$ 
is an eigenstate of the electronic problem which is parameterized
in terms of $\delta_p$. Therefore, the minimization of the force
fields has to be done self-consistently.\cite{Bursill2003}
{\begin{enumerate}\renewcommand{\theenumi}{\roman{enumi}}
\item In step $k$ of the iteration ($k=1,2,\ldots $),
the target eigenstate $|\Psi_k\rangle$, e.g., $G_k$, $S_k$, or $T_k$,
is calculated for $\delta_{k;l}$ using the 
infinite-lattice DMRG algorithm.
In all our cases, the initial choice $\delta_{1;l}=0$
for $l=1,\ldots,N-1$ leads to converged solutions.
\item For given~$k$ and fixed quantum-mechanical expectation values 
in $|\Psi_k\rangle$, the distortion energies are determined
iteratively. 

To this end, the condition $f_l=0$ in~(\ref{eq:def-f}) is used
to determine the distortion energies for the next iteration,
$\delta_{k;l}^{n+1}=2\pi t_0 \lambda_l(-\Gamma_k^{n}
+F_l[\delta_{k;p}^{n}])$,
($n\geq 0$, $\delta_{k;l}^{0}=\delta_{k;l}$). Here, the Lagrange parameter 
follows from~(\ref{eq:Gamma-condition}) 
as $\Gamma_k^{n}=\sum_l \lambda_l F_l[\delta_{k;p}^{n}]/\sum_l \lambda_l$.

The distortion energies typically converge after some five to fifteen 
iterations.
The converged solution defines $\delta_{k+1;l}=\lim_{n\to\infty}\delta_{k;l}^n$
for the next iteration in~$k$.
\item The steps (i) and~(ii) are repeated until a
converged set of distortion energies and DMRG energies
for the states
are obtained, $\delta_l=\lim_{k\to\infty}\delta_{k;l}$,
$|\Psi\rangle=\lim_{k\to\infty}|\Psi_k\rangle$.
\end{enumerate}}
Strictly speaking, the condition of a fixed chain length ${\ell}_c$
corresponds to $x_{1,N}^2+y_{1,N}^2={\ell}_{c}^2$. We have verified
numerically that the condition~(\ref{eq:Gamma-condition})
preserves the chain length up to 0.1\% for $N\lesssim 100$.

\subsubsection{Polaronic energies}

It is important to note that we optimize 
the lattice structure and the corresponding distortion 
energies for each state separately.
Our excited states contain all polaronic energy contributions, i.e., 
we give their {\em relaxed\/} energies.
This polaronic
relaxation was not taken into account in our previous study.\cite{all-of-us} 
There, we studied {\em rigid-lattice\/} transitions 
with fixed electron transfer amplitudes
$t_s$, $t_d$, and $t_t$, which correspond 
to the Lewis structure of Fig.~\ref{fig:structure}.
In general, they are higher in energy than the corresponding {\em relaxed\/} 
excitations. 

It is not a priori clear whether the relaxed or the
rigid-lattice energies should be compared to experiment.
For the optical singlet excitation, no Stokes shift is observed
between absorption and fluorescence spectra~\cite{Schott-review} so that,
in the Franck--Condon picture, 
the exciton creation process corresponds to a vertical transition.

In order to estimate the polaronic contribution to the energy,
we calculate the energy of excited states in
the rigid-lattice approximation for our optimal parameter set, 
see Sect.~\ref{sec:results}.
For $(t_0^*=2.4\, {\rm eV}, \alpha^*=3.4\, {\rm eV}/\hbox{\AA},
U^*=6\, {\rm eV}, V^*=3\, {\rm eV})$,
the relaxed energy of the single-particle gap is
$E_{\rm gap}^{\rm relaxed}=2.45\, {\rm eV}$ whereas the energy of
a single-particle excitation with fixed electron transfer
matrix elements $t_{\rm s}^{\rm G}$, $t_{\rm d}^{\rm G}$, and $t_{\rm t}^{\rm G}$
leads to
$E_{\rm gap}^{\rm rigid}=2.47\, {\rm eV}$, see table~\ref{tab:BCMU-energies}.
Thus, the energy relaxation due to the polaron formation
amounts to about
$\delta_{\rm polaron}=0.02\, {\rm eV}$. 
In the band picture, the singlet exciton is a bound state of 
particle-hole excitations. Correspondingly, the polaronic shift
in $E_{\rm S}$ should be about twice as large as $\delta_{\rm polaron}$,
as indeed observed,
$E_{\rm S}^{\rm relaxed}=2.00\, {\rm eV}$ and
$E_{\rm S}^{\rm rigid}=2.05\, {\rm eV}$,
so that
$E_{\rm S}^{\rm rigid}-E_{\rm S}^{\rm relaxed}=0.05\, {\rm eV}\approx
2\delta_{\rm polaron}$. 

The same amount of polaronic relaxation energy is
observed for the singlet-triplet gap,
$E_{\rm T}^{\rm rigid}-E_{\rm T}^{\rm relaxed}=0.06\, {\rm eV}$.
When we start from the relaxed triplet ground state, we find
for the optically excited state~T$^*$ that
$E_{{\rm T}^*}^{\rm rigid}=1.28\, {\rm eV}$ 
whereas
$E_{{\rm T}^*}^{\rm relaxed}=1.25\, {\rm eV}$.
It is seen that the polaronic relaxation energy amounts to about
$\delta_{\rm polaron}$ also in the triplet sector.

Our observation of a fairly small polaronic relaxation energy
ties in with the fact that our electron-lattice
coupling is small, $\lambda_l< 0.1$, and the adiabatic approximation
is valid. In our comparison with experiment below, we show the energies
for transitions between lattice-relaxed configurations.

\subsection{Single-particle gap and in-gap excitations}
\label{subsec:energies}

The band-gap or single-particle gap $E_{\rm gap}$ is defined by the difference
in chemical potentials for a system with $N_{\rm e}$ and $N_{\rm e}-1$ particles,
\begin{eqnarray}
E_{\rm gap} &=& \mu(N_{\rm e})-\mu(N_{\rm e}-1) \; , \nonumber \\
\mu(N_{\rm e})&=& E_{\rm G}(N_{\rm e}+1)-E_{\rm G}(N_{\rm e})\; ,
\end{eqnarray}
where $E_{\rm G}(N_{\rm e})$ is the energy of the $N_{\rm e}$-particle ground 
state~G.
In the presence of particle-hole symmetry at half band-filling, we have
\begin{equation}
E_{\rm gap} = 2\mu(N_{\rm e})
\end{equation}
for the minimal
energy of a single-particle excitation.

In poly-diacetylenes, the singlet exciton and its vibronic replicas
carry most of the oscillator strength of the optical excitations.
The quadratic Stark effect in the electro-absorption proves that
they are bound states of electron-hole excitations.~\cite{Weiser}
The exciton energy thus defines the optical gap,
\begin{equation}
\Delta_{\rm opt}^{\rm s}= E_{\rm S}(N_{\rm e}=N)-E_{\rm G}(N_{\rm e}=N) \;, 
\end{equation}
where $E_{\rm S}(N_{\rm e}=N)$ is the energy of the first excited state of the
half-filled system with symmetry $B_u$.
The binding energy of the exciton is then obtained as
\begin{equation}
\Delta_{\rm ex}^{\rm s} = E_{\rm gap}- E_{\rm S} \; .
\end{equation}
Note that we calculate for finite-size systems so that all quantities
must be extrapolated into the thermodynamic limit, $N\to \infty$.

For a full account of all in-gap states, we target up to five
states simultaneously in the spin-singlet and spin-triplet sectors, 
respectively.
Note that the lattice relaxation must be done for each state separately.
These calculations represent the most time consuming part of our
investigations.

\subsection{Numerical procedure}
\label{Subsec:DMRG}

In this work we present results from numerical density-matrix 
renormalization-group (DMRG)~\cite{steve} calculations 
on finite chains with open boundary condition (OBC) 
using an adopted version used previously.\cite{all-of-us} 
The discarded weight was kept below $\eta=10^{-6}$ 
for all calculations by employing the dynamical block-state selection
(DBSS) procedure.\cite{dbss-1,dbss-2} 
We have set the minimum number of block states to 
$M_{\rm min}=400$ and used three sweeps. The maximum value of the number of 
block states varied around $M_{\rm max}=600$. As benchmarks we  
compared our DMRG energies for some selected parameter values
with those from a DMRG code used 
earlier by Race, Barford, and Bursill.\cite{Bursill2003} 
The latter one, however, only uses
the infinite-lattice procedure so that 
our variational energies are always slightly lower. 

The ground state as well as all excited states have been 
targeted and relaxed individually using $k=5$ to $15$ relaxation iteration 
steps to reach the pre-set convergence criterion on $\delta_l$. 
Note that each relaxation iteration step requires 
a full DMRG run with three sweeps. 
The whole relaxation procedure has been performed for all target states 
and for all chain lengths independently from $N=6$ up to $N=66$, 
in steps of $\Delta N=4$. For our optimized parameter set we have performed 
calculations for up to $N=102$ sites.

\begin{figure}[tb]
\includegraphics[scale=0.43]{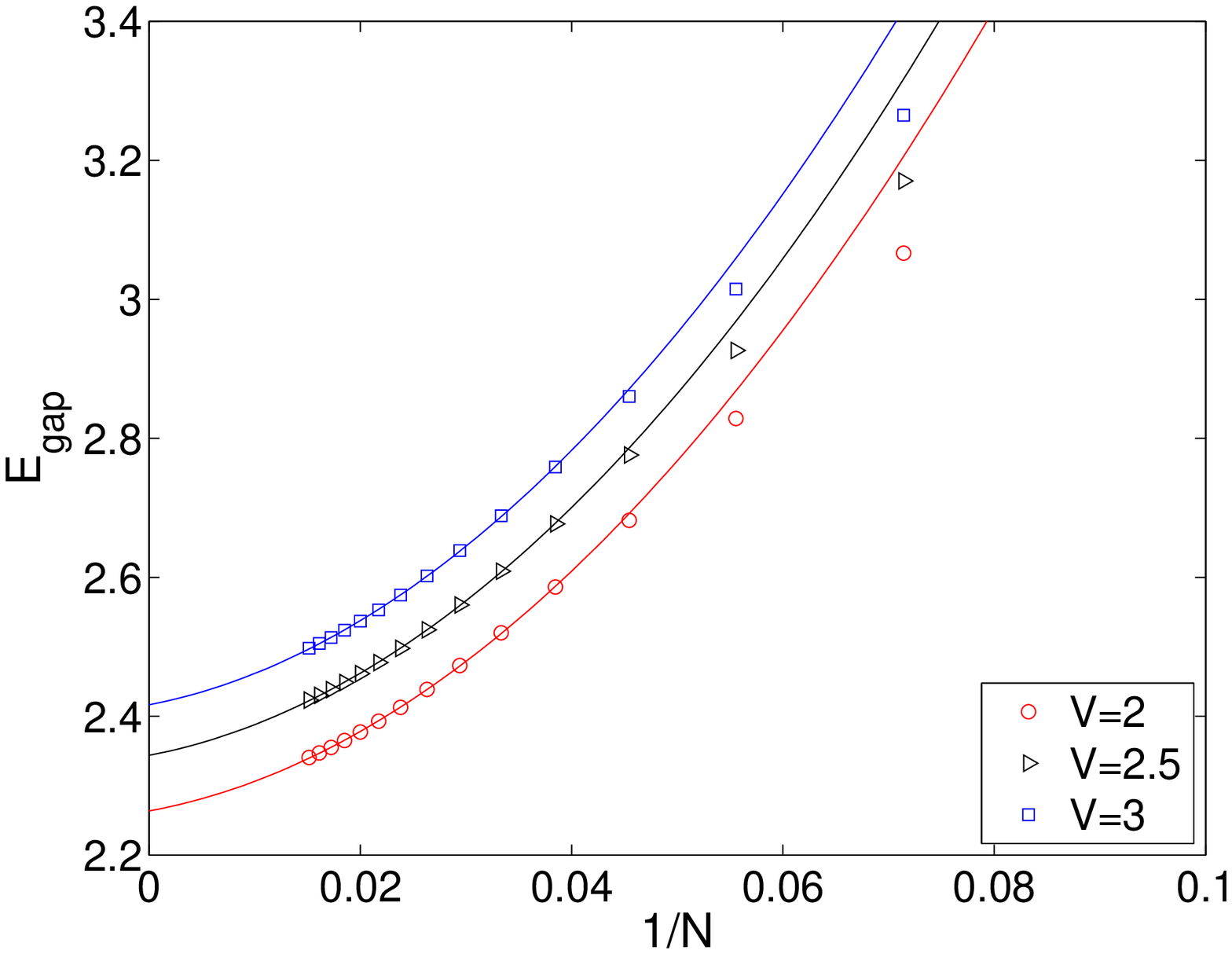}
\caption{(Color online) Finite-size scaling of the charge gap $E_{\rm gap}$,
calculated for $t_0=2.4\, {\rm eV}$, 
$\alpha=3.4\, {\rm eV}/\hbox{\AA}$, $U=6\, {\rm eV}$, 
$V=2.0,2.5,3.0\, {\rm eV}$,
and $14\leq N\leq 66$ sites. The lines are quadratic fits. 
\label{fig:singletsector-finite}}

\vspace*{3pt}

\includegraphics[scale=0.43]{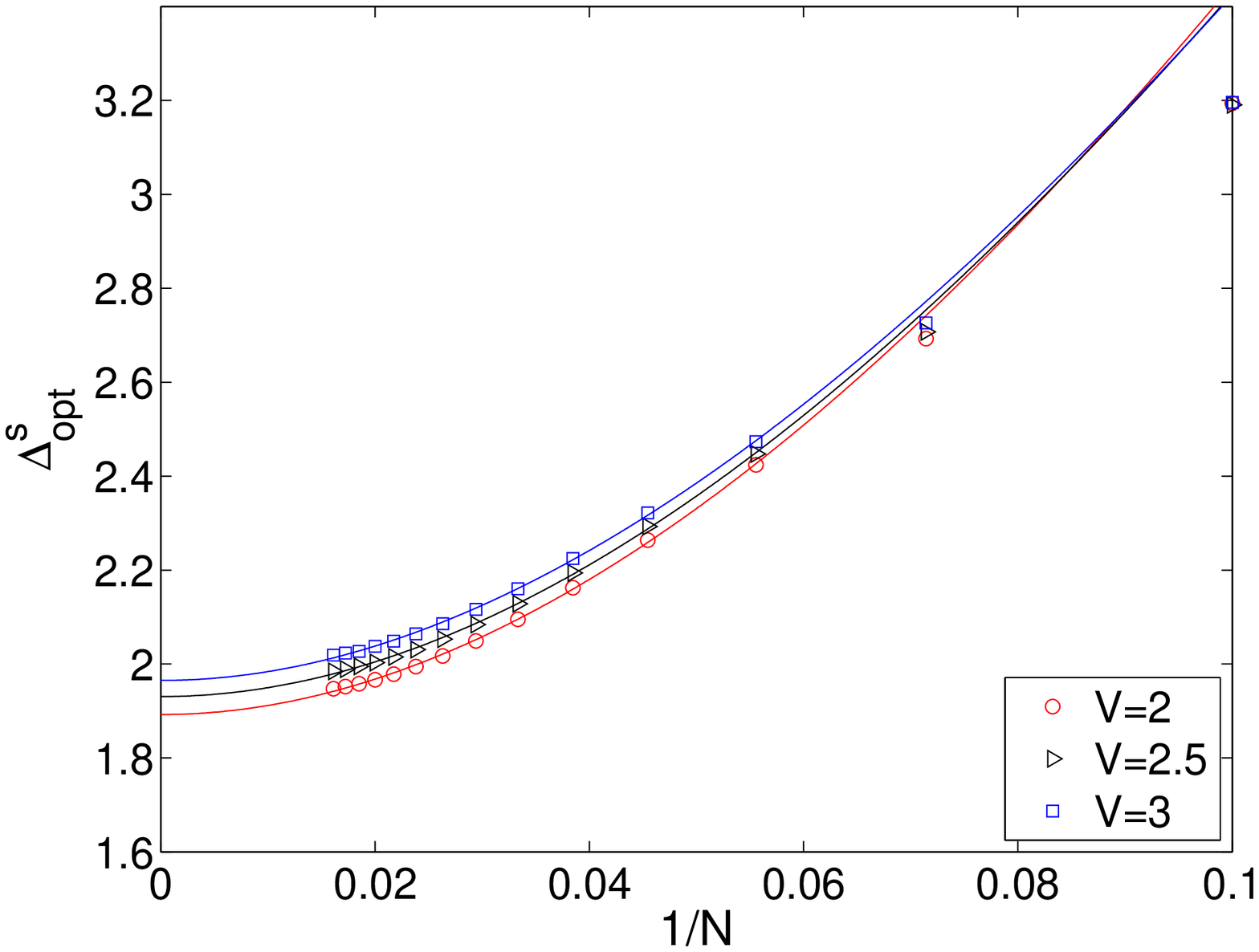}
\caption{(Color online) Finite-size scaling 
of the singlet exciton energy $E_{\rm S}$ 
(optical gap $\Delta_{\rm opt}^{\rm s}$), 
calculated for $t_0=2.4\, {\rm eV}$, 
$\alpha=3.4\, {\rm eV}/\hbox{\AA}$, 
$U=6\, {\rm eV}$, 
$V=2.0,2.5,3.0\, {\rm eV}$,
and $10\leq N\leq 66$ sites. The lines are quadratic fits. 
\label{fig:singletsector-finite2}}
\end{figure}

The spin multiplicity of the converged target states was calculated from
the expectation value of the operator for the total spin $\hat{S}^2$, employing
the expectation values of the corresponding correlation 
functions.\cite{barcza}
In order to determine the optically dark in-gap singlet states 
we have shifted the energies of the triplet states out of the gap
by adding the term $\sum_{ij} S^-_i S^+_j$ to the Hamiltonian.
As an alternative procedure, we have also identified 
the exciton states by calculating 
the dipole strength as introduced in Ref.~[\onlinecite{all-of-us}] 
in which 
the reduced density matrix of the target state was constructed 
from the reduced density matrices of the ten lowest eigenstates.

For the 81~parameter points of the four-dimensional search 
space ($t_0, \alpha, U, V)$ and for the energies of the states 
($G, S, E_{\rm gap}, T, T^{*}, X_l, Y_l$) 
shown in Fig.~\ref{fig:energylevels}, we have taken
on average $k=10$ relaxation steps for the 16 to 24~different chain 
lengths for single and multiple target states which results in
about 130.000~full DMRG runs. We estimate that 
the calculations consumed overall about
45~CPU years which were provided by 100~parallelized CPUs so that
the calculations took some five months in real time. 

The PDAs are charge and spin insulators, i.e., the gaps for single-particle, 
optical, and magnetic excitations are finite. The materials are characterized
by finite correlation lengths. Therefore, end effects decay 
exponentially, and local operators that are calculated in the middle
of the chain display a regular behavior as a function of inverse system size.
Thus, various quantities that we calculate for finite chain lengths~$N$
can be extrapolated reliably to the thermodynamic limit, $N\to\infty$, 
by using a second-order polynomial fit.

As an example, in Fig.~\ref{fig:singletsector-finite} 
we show the charge gap calculated for $t_0=2.4$, 
$\alpha=3.4\, {\rm eV}/\hbox{\AA}$, $U=6\, {\rm eV}$,
and $V=2.0,2.5, 3.0\, {\rm eV}$,
as a function of $1/N$. 
As seen from the figure,
the second-order polynomial fits permit 
a reliable extrapolation but the quadratic curvature becomes dominant 
for $N\gtrsim 50$ only. Therefore, a study of long chains is mandatory.
The appearance of the inflection point at sizable chain lengths
is more pronounced for the gap states than for the single-particle gap. 
An example, the optical gap, is shown 
in Fig.~\ref{fig:singletsector-finite2}.

\begin{figure}[hb]
\includegraphics[scale=0.43]{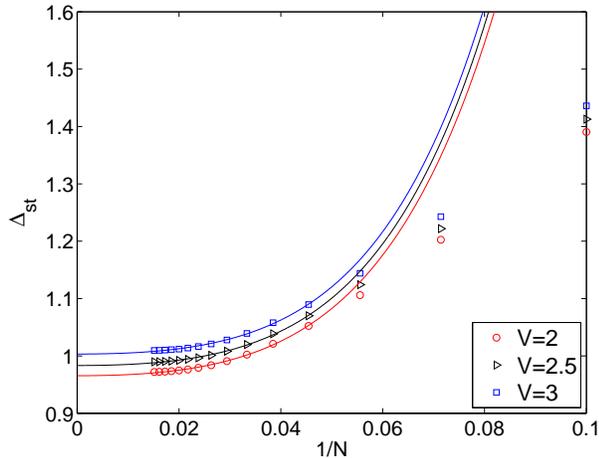}
\caption{(Color online) Finite-size scaling of the triplet ground-state energy
$E_{\rm T}$ (singlet-triplet gap $\Delta_{\rm st}$), 
calculated for $t_0=2.4\, {\rm eV}$, 
$\alpha=3.4\, {\rm eV}/\hbox{\AA}$, $U=6\, {\rm eV}$, 
$V=2.0,2.5,3.0\, {\rm eV}$,
and $10\leq N\leq 66$ sites. The lines are quadratic fits. 
\label{fig:tripletsector-finite}}
\end{figure}

\mbox{}

\begin{figure}[hb]
\includegraphics[scale=0.43]{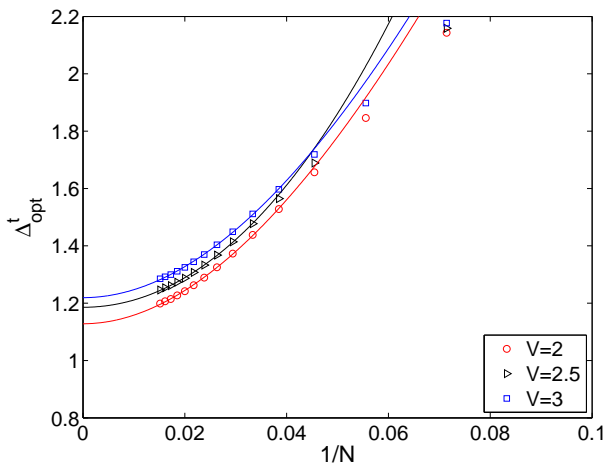}
\caption{(Color online) Finite-size scaling of 
the optical gap in the triplet sector 
$\Delta_{\rm opt}^{\rm t}=E_{\rm T^*}-E_{\rm T}$, 
calculated for $t_0=2.4\, {\rm eV}$, 
$\alpha=3.4\, {\rm eV}/\hbox{\AA}$, 
$U=6\, {\rm eV}$, 
$V=2.0,2.5,3.0\, {\rm eV}$,
and $14\leq N\leq 66$ sites. The lines are quadratic fits. 
\label{fig:tripletsector-finite2}}
\end{figure}

In Figs.~\ref{fig:tripletsector-finite} and~\ref{fig:tripletsector-finite2}
we show finite-size results for the triplet ground-state energy~$E_{\rm T}$
(singlet-triplet gap $\Delta_{\rm st}$) and
for the optical gap in the triplet sector 
$\Delta_{\rm opt}^{\rm t}=E_{\rm T^*}-E_{\rm T}$, respectively, 
for $t_0=2.4\, {\rm eV}$, 
$\alpha=3.4\, {\rm eV}/\hbox{\AA}$, $U=6\, {\rm eV}$, 
$V=2.0,2.5,3.0\, {\rm eV}$.
The ground-state energy $E_{\rm T}$ 
in the triplet sector rapidly converges as a function of 
inverse system size~$1/N$ because the state is deep in the gap.
The energy $E_{\rm T^*}$ of its optical excitation~T$^*$ is close
to the threshold $E_{\rm gap}$ for single-particle excitations
so that long chains must be studied for a reliable extrapolation
to the thermodynamic limit.
For the scan of our parameter regime as specified in the next section, 
we limit ourselves to chains of length $N\leq 66$.
The accuracy of the extrapolation 
is better than $\delta E=0.05\, {\rm eV}$.

\begin{figure}[h]
\includegraphics[scale=0.42]{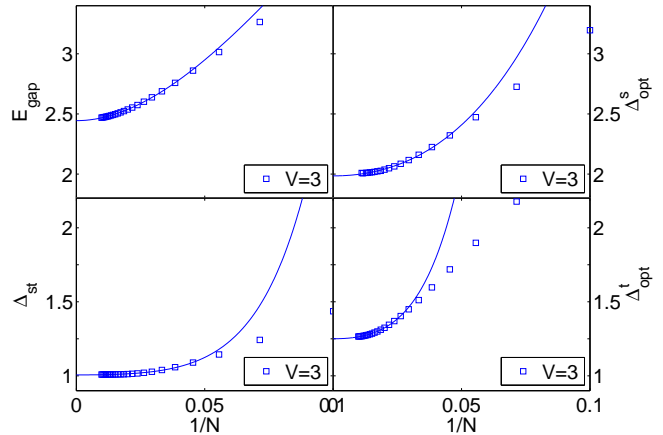}
\caption{(Color online) 
Finite-size scaling of 
(a)~the charge gap $E_{\rm gap}$, (b)~the optical gap $\Delta_{\rm opt}^{\rm s}$,
(c)~the singlet triplet gap~$\Delta_{\rm st}$, and (d)~the optical gap
in the triplet sector~$\Delta_{\rm opt}^{\rm t}$,
calculated for the optimal parameter set
($t_0^*=2.4\, {\rm eV}$, 
$\alpha^*=3.4\, {\rm eV}/\hbox{\AA}$, $U^*=6\, {\rm eV}$, 
$V^*=3\, {\rm eV}$),
and $N\leq 102$ sites. The lines are quadratic fits. 
The extrapolated values are summarized in Tab.~\ref{tab:BCMU-energies}.
\label{fig:favorites-finite}}
\end{figure}

Before we detail our optimization of the parameter set
in Sect.~\ref{subsec:optimization}, we show in Fig.~\ref{fig:favorites-finite}
the finite-size scaling of all four gaps for our 
optimal choice of parameters
$(t_0^*=2.4\, {\rm eV}, \alpha^*=3.4\, {\rm eV}/\hbox{\AA},
U^*=6\, {\rm eV}, V^*=3\, {\rm eV})$ for which we investigate chains
with up to $N=102$ sites. 
As can be seen from Fig.~\ref{fig:favorites-finite},
the finite-size extrapolation can be done very accurately with
an uncertainty of $\delta E\lesssim 0.01\, {\rm eV}$.
The extrapolated values are
$E_{\rm gap}=2.45\, {\rm eV}$, $\Delta_{\rm opt}^{\rm s}=2.00\, {\rm eV}$,
$\Delta_{\rm st}=1.00\, {\rm eV}$, and $\Delta_{\rm opt}^{\rm t}=1.25\, {\rm eV}$,
see also table~\ref{tab:BCMU-energies}.

\section{Parameter regime}
\label{sec:guessparameters}

Before we present our results in Sect.~\ref{sec:results}, 
we give arguments for the parameter regime
used in our study, as summarized in table~\ref{tab:parameters}.

\subsection{Spring constants}
\label{subsec:springconstants}

At first sight, it seems to be easy to obtain the spring
constants~$K$ and~$G$ because they can be inferred from Raman scattering
data for short molecules. Unfortunately, the influence of the
delocalized $\pi$~electrons appears to be crucial.
For example, early Raman experiments in ethane,~\cite{SutherlandDennison} 
H$_3$C$-$CH$_3$, give
$K_{\sigma}^{\rm e}=31\, {\rm eV}/\mbox{\AA}{}^2$
whereas for the single-bond in diacetylene,~\cite{VallanceJones} 
HC$\equiv$C$-$C$\equiv$CH, 
$K_{\sigma}^{\rm da}=45\, {\rm eV}/\mbox{\AA}{}^2$
is found. In polymers, the situation is equally ambiguous because
the theoretical analysis of the same Raman data for poly-acetylene
leads to the same set of concurring values~\cite{Girlandoetal} 
$K_{\sigma}^{\rm PA,1}=31\, {\rm eV}/\mbox{\AA}{}^2$
and 
$K_{\sigma}^{\rm PA,2}=46\, {\rm eV}/\mbox{\AA}{}^2$.~\cite{Ehrenfreund}
In  the present work, we investigate the consequences of
a strong spring constant for the $\sigma$-bond, 
$K> 40 \, {\rm eV}/\mbox{\AA}{}^2$. We plan to present
a detailed study of the vibrational properties in the near future.

In order to estimate~$K$ (and~$G$), we calculate the optical phonon
spectrum from a simple classical model.
The carbon atoms of mass~$M=12 u$ in the Lewis
structure of Fig.~\ref{fig:structure}
are linked by spring constants of strength~$K$ and~$G=gK$
in the sequence $(G,K,K,K)$ in the unit cell, corresponding
to the $\sigma$-$p_y$-bond and the three $\sigma$-bonds.
The optical phonons for two-dimensional vibrations of the chain
are derived in App.~\ref{app:phonons}. 
There are four positive solutions of the characteristic equation
for the phonon frequencies, $\omega_a< \omega_b=\sqrt{2K/M}< \omega_c<\omega_d$,
where $\omega_b$ is the resonance frequency of two carbon atoms
linked by the spring constant~$K$.~\cite{Ehrenfreund}
The frequencies $\omega_{a,c,d}$ are obtained from the zeros of the
third-order polynomial
\begin{equation}
p(y)=y^3-2(g+2)y^2+(7/2+6g)y-3g
\end{equation}
as $\omega_{a,c,d}=\sqrt{y_{a,c,d}K/M}$, $p(y_{a,c,d})=0$.

In comparison with experiment, see Sect.~\ref{subsec:phononreplicas}, we
assign $\omega_d(g)=\omega_{\rm T}$ and $\omega_c(g)=\omega_{\rm D}$. From
$(\omega_{\rm T}/\omega_{\rm D})^2=(0.261/0.181)^2=2.079$
we find $g_0=1.547$, $y_d(g_0)=4.463$,
$y_c(g_0)=2.147$, and $y_a(g_0)=0.4845$. From 
$y_c(g_0)=M \omega_{\rm D}^2/K$
we obtain $K=44.1\, {\rm eV}/\hbox{\AA}{}^2$ which
agrees with results obtained for poly-acetylene chains,~\cite{Ehrenfreund}
$K_{\sigma}^{\rm PA,2}=46\, {\rm eV}/\hbox{\AA}{}^2$. In addition,
we find $G=g K=68.3\, {\rm eV}/\hbox{\AA}{}^2$ for the $\sigma$-$p_y$-bond.
For comparison, the spring constant in ethene (ethylene) 
was derived as 
$K_{\sigma-p_y}^{\rm eth}=60 \, {\rm eV}/\mbox{\AA}{}^2$.~\cite{Herzberg}

The other two optical phonons have the energies
$\hbar\omega_b
=\hbar\sqrt{2K/M}=0.175\, {\rm eV}$,
and
$\hbar\omega_a
=\hbar\sqrt{K/M}\sqrt{y_a(g_0)}=0.086\, {\rm eV}$,
respectively.
These values are
in good agreement with experiment,
$\hbar\omega_b\approx \hbar\omega_{\rm D^*}=0.155\, {\rm eV}$,
and
$\hbar\omega_a\approx \hbar\omega_{\rm S}=0.090\, {\rm eV}$,
see Sect.~\ref{subsec:phononreplicas}.
Note that we use experimental data for comparison which
include the influence of the $\pi$-electrons whereas
for our model calculations we employ bare values for the backbone.
The influence of the itinerant $\pi$ electrons 
must be calculated self-consistently so that
the values for~$K$ and~$G$ need further refinement.
This task is left for a future study.

\subsection{Electron-phonon coupling}

Next, we discuss the bare bandstructure for non-interacting electrons
and 
estimate the size of the electron-phonon coupling constant~$\alpha$.

\subsubsection{Bare bandstructure}

The bare bandstructure for the ground state
with filled valence bands with energies
$E_{v,2}(k)=-\epsilon_2(k)$
and $E_{v,1}(k)=-\epsilon_1(k)$
and empty conduction bands with energies
$E_{c,1}(k)=\epsilon_1(k)$
and $E_{c,2}(k)=\epsilon_2(k)$ is derived in App.~\ref{app:barebands},
see eq.~(\ref{appeq:eps}).
The bare gap is given by $\Delta^{\rm bare}=2 \epsilon_1(k=0)$.
The Coulomb interaction enhances all gaps~\cite{Rissler-Grage} so that the
singlet-triplet gap $\Delta_{\rm st}=E_{\rm T}-E_{\rm G}\approx 1\, {\rm eV}$
will be larger than $\Delta^{\rm bare}$.
Therefore, the size of the bare band-gap constrains the possible
values for the electron-phonon coupling.

In the absence of Coulomb interactions,
the values for $t_{\rm s}$, $t_{\rm d}$, and $t_{\rm t}$ 
must be determined from the minimization
of the total ground-state energy per unit cell,
\begin{eqnarray}
e_{\rm tot}(\delta_{\rm d},\delta_t)
&=& e_{\rm kin}(\delta_{\rm d},\delta_t) + e_{\rm pot}(\delta_{\rm d},\delta_t) 
\nonumber \; , \\
e_{\rm pot}(\delta_{\rm d},\delta_t) &=&
\frac{1}{4\pi t_0\lambda} 
\left(\delta_{\rm d}^2
+g \delta_t^2+\frac{1}{2}(\delta_t+\delta_{\rm d})^2\right) \; ,
\label{eq:minimize}
\end{eqnarray}
where $g=G/K$ and $\lambda=2\alpha^2/(\pi t_0 K)$.
The kinetic energy of the electrons is given by
\begin{equation}
e_{\rm kin}(\delta_{\rm d},\delta_t)=
-2\int_{-\pi}^{\pi} \frac{{\rm d}k}{2\pi} 
\left(\epsilon_1(k)+\epsilon_2(k)\right) \; ,
\end{equation}
where the factor two accounts for the spin degeneracy.

The electron transfer matrix elements $t_{\rm s,d,t}$
and the distortion corrections $\delta_{\rm s,d,t}$ are related by
$t_{\rm s}= t_0 -(\delta_{\rm t}+\delta_{\rm d})/4$, 
$t_{\rm d}= t_0 +\delta_{\rm d}/2$, and 
$t_{\rm t}= t_0 +\delta_{\rm t}/2 + \delta^{\rm e}/2$. 
Here, we used the fact that $2\delta_{\rm s}+\delta_{\rm d}+\delta_{\rm t}=0$ 
because the length of the unit cell is not changed by 
the intrinsic distortion.
The strength of the extrinsic dimerization~$\delta^{\rm e}$
follows from the solution of eq.~(\ref{appeq:implicit}) 
of App.~\ref{appA} for $U=V=0$,
\begin{equation}
\frac{\delta^{\rm e}(U=V=0)}{\pi t_0\lambda}
= - \frac{8 T_1(\delta^{\rm e})}{2\epsilon_0(\delta^{\rm e})}
= \frac{8T_1(\delta^{\rm e})}{4 T_1(\delta^{\rm e})}=2
\end{equation}
so that 
$\delta^{\rm e}(U=V=0)= 4\alpha^2/K$.

The numerical minimization of $e_{\rm tot}(\delta_{\rm d},\delta_{\rm t})$, 
eq.~(\ref{eq:minimize}),
leads to the somewhat surprising result that 
the bare gap $\Delta^{\rm bare}(t_0,\alpha)$
very weakly depends on~$t_0$. For $\alpha=3.5\, {\rm eV}/\hbox{\AA}$,
$K= 44\, {\rm eV}/\hbox{\AA}{}^2$, and
$G= 68\, {\rm eV}/\hbox{\AA}{}^2$,
we find that $\Delta^{\rm bare}(2,3.5)=0.878\, {\rm eV}$
and $\Delta^{\rm bare}(2.4,3.5)=0.835\, {\rm eV}$;
it even decreases slightly with increasing~$t_0$.
In contrast, the bare band-gap strongly increases as a function of $\alpha$.
For $t_0=2.4\, {\rm eV}$, $K= 44\, {\rm eV}/\hbox{\AA}{}^2$, and
$G= 68\, {\rm eV}/\hbox{\AA}{}^2$,
we find
$\Delta^{\rm bare}(2.4,3.4)=0.776\, {\rm eV}$
and
$\Delta^{\rm bare}(2.4,3.6)=0.897\, {\rm eV}$.
The bare band-gap becomes larger than $\Delta_{\rm st}=1\, {\rm eV}$
for $t_0=2.4\, {\rm eV}$ and $\alpha=3.8\, {\rm eV}/\mbox{\AA}$. 
Therefore, we must use smaller values for $\alpha$
as derived and used previously.~\cite{Ehrenfreund,Bursill2003}

\subsubsection{Intrinsic and extrinsic Peierls distortion}

In poly-acetylene (PA), the mobile $\pi$-electrons dimerize the
chain. This intrinsic Peierls effect results in a measured bond length
alternation of $\Delta r=0.04\, \mbox{\AA}$, i.e.,
long and short bonds of length $r_{\rm s}^{\rm PA}=1.44\, \mbox{\AA}$
and $r_{\rm d}^{\rm PA}=1.36\, \mbox{\AA}$ alternate along the 
chain.~\cite{PAdistortion}
Almost the same amount of alternation in $r_{\rm s}$ and $r_{\rm d}$
is seen in PDAs, see Sect.~\ref{subsubsec:latticeparameters}.
Previous DMRG studies~\cite{Bursill2003} lead to $\Delta r=0.03\, \mbox{\AA}$.

When we assume that the intrinsic Peierls effect
affects the triple bond in the same way as the double bond,
we come to the conclusion that the length~$R_1$ of the $\sigma$-$p_y$-bond
before the intrinsic dimerization is $R_1\approx 1.25\, \mbox{\AA}$
assuming $r_{\rm t}=1.21\, \mbox{\AA}$.
Therefore, the extrinsic dimerization due to the $p_y$-bond accounts
for $r_0-R_1=\delta^{\rm e}/(2\alpha)=0.15\, \mbox{\AA}$.
For non-interacting electrons, we have $\delta^{\rm e}(U=V=0)=4\alpha^2/K$,
independent of~$t_0$. Thus,
we arrive at the estimate $\alpha\approx 0.15\hbox{\AA} (K/2)$ which 
gives $\alpha\approx 3.3\, {\rm eV}/\hbox{\AA}$ for 
$K=44\, {\rm eV}/\hbox{\AA}{}^2$.

\begin{figure}[tb]
\includegraphics[scale=0.40]{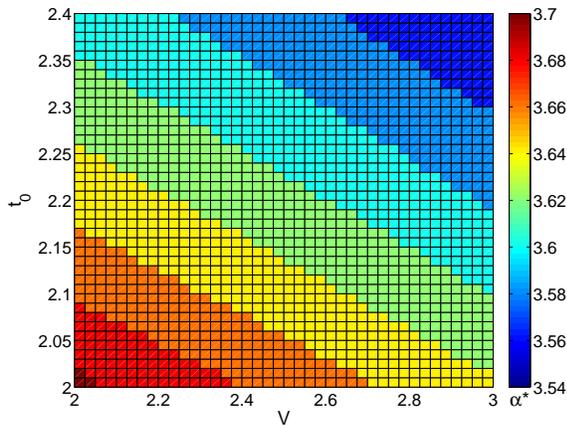}
\caption{(Color online) Regions in the parameter space $t_0$-$V$
which lead to $R_1=1.25\, \mbox{\AA}$ for $U=6\, {\rm eV}$. 
The color coding gives the appropriate value for $\alpha$. 
\label{fig:findalpha}}
\end{figure}

Even in the presence of Coulomb interactions, 
the calculation of $\delta^{\rm e}$ is simple because
it requires the solution of a two-site
problem only, see App.~\ref{appA}. Therefore, a complete parameter scan
is readily accomplished. As we shall argue below, the on-site Coulomb
repulsion~$U$ is quite substantial. Therefore,
in Fig.~\ref{fig:findalpha} we show parameter regions 
in ($t_0,\alpha,V$) for $U=6\, {\rm eV}$ 
that correspond to $\delta^{\rm e}/(2\alpha)=0.15\, \mbox{\AA}$. 
As compared to non-interacting electrons,
the electron-phonon coupling
$\alpha$ has to be increased by some 10\% to generate the same
extrinsic Peierls dimerization for the interacting two-site system. 
This indicates that 
the Coulomb interaction makes the bonds noticeably
stiffer. In order to account for
this effect, our spring constant~$K$ from Sect.~\ref{subsec:springconstants}
should be reduced by at least ten percent;
a more thorough scan for the $K$-parameter will be done in a future study.

Fig.~\ref{fig:findalpha} shows that
the dependence on $t_0$ and~$V$ is rather weak: it is mostly
the Peierls coupling~$\alpha$ that determines the size of the bond length
shift. From the data in Fig.~\ref{fig:findalpha} we conclude that
$\alpha\approx (3.5\pm 0.1)\, {\rm eV}/\mbox{\AA}$ 
is a reasonable starting point, whereby we compensate 
our somewhat too large spring constant~$K$.

\subsection{Coulomb parameters}

Previous studies~\cite{LouieRohlfing,LDA-GW-BSE,Bursill2003,all-of-us}
succeeded to reproduce the charge gap and the energy of the
singlet exciton. However, the complexity of the in-gap states
could not be recovered. In particular, the splitting $\Delta_{\rm st}$
of the singlet and triplet ground states
cannot be reproduced as it comes out substantially too high. 
Moreover, in previous approaches 
no dark singlet states have been found that lie
energetically below the singlet exciton. 

The energetic positions of the states~T and X$_{1,2}$ are a clear signal
of substantial electronic correlations induced by the Hubbard interaction~$U$.
Our initial calculations with small ratios $U/V$ put~T and X$_{1,2}$ too
high in energy as compared to experiment.\cite{Barcza-unpublished} 
As we shall show in the next section, we find a reasonably good
description of the level scheme in Fig.~\ref{fig:energylevels}
only for substantial~$U$ and comparably small~$V$,
$5\, {\rm eV} \leq U \leq 6\, {\rm eV}$ and
$2\, {\rm eV} \leq V \leq 3\, {\rm eV}$, as indicated in 
table~\ref{tab:parameters}, so that $U/V=\kappa \approx 2$ holds.
Note that we included the dielectric constant~$\epsilon_d$ explicitly
in the Ohno potential~(\ref{eq:Ohno-potential})
because we treat chains immersed in their monomer matrix.

Substantial values for the Coulomb interaction were advanced
by Chandross and Mazumdar~\cite{Chandross} as a result of their model study
of poly-phenylene-vinylene (PPV) thin films.
In order to describe the linear and non-linear optical properties
of PPV, they proposed $U_{\rm CM}=8\, {\rm eV}$
with $U_{\rm CM}/V_{\rm CM}=\kappa =2$.\cite{Chandross,Mazumdar-Wang-Zhao}
Note, however, that these authors worked with fixed lattice parameters,
i.e., without lattice relaxations for the excitations,
and employed an approximation (single configuration interaction, SCI)
to calculate optical excitations.

The bare bandwidth in our calculations is $W\approx 4t_0\gtrsim 9\, {\rm eV}$ 
which still is larger than the on-site interaction. Therefore,
the system is still far from the spin-Peierls limit.
At the same time, however, the correlations are strong enough to
impede weak-coupling approaches.\cite{Rissler-Grage}

\section{Results}
\label{sec:results}

First, we scan our parameter space and determine
our best parameter set ($t_0, \alpha, U, V$).
Next, we analyze the energy levels of optically dark in-gap states
and comment on the lattice parameters.

\subsection{Optimization of the parameter set}
\label{subsec:optimization}

For each choice of the parameter set 
($t_0, \alpha, U, V$),
we calculate the ground-state energies at half band-filling
and one additional particle,
and the energy of the three excited state~S, T, and T$^*$
at half band-filling
from which we determine the four gaps $E_{\rm gap}$, $\Delta_{\rm opt}^{\rm s}$,
$\Delta_{\rm st}$, and $\Delta_{\rm opt}^{\rm t}$
for systems with size $10\leq N\leq 66$. 
The lattice geometry of all states is relaxed, see Sect.~\ref{sec:method}.
We performed calculations for ($t_0,\alpha,U,V)$
for a broad range of parameters using small system
sizes up to~$N=30$ sites. For the optimal range 
$(t_0=2.0,2.2,2.3,2.4\, {\rm eV}$,
$\alpha=3.4,3.5,3.6\, {\rm eV}/\hbox{\AA}$,
$U=5.0,5.5,6.0\, {\rm eV}$,
$V=2.0,2.5,3.0\, {\rm eV}$) we investigated systems with up to 
$N=66$~sites, i.e., we address altogether 108~different parameter sets.

\begin{figure}[hb]
\includegraphics[scale=0.42]{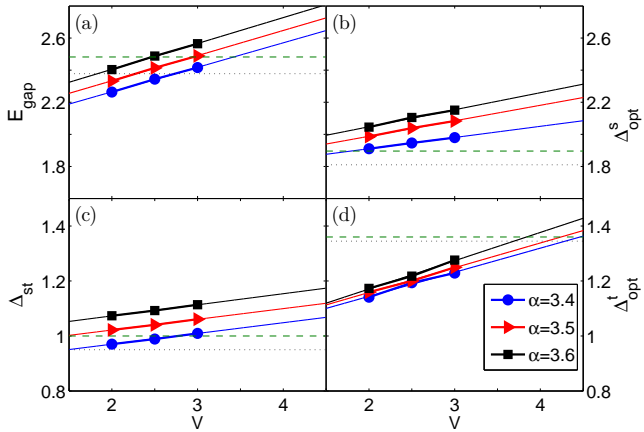}
\caption{(Color online) Extrapolated energies as a function
of~$V$ for $\alpha=3.4,3.5,3.6\, {\rm eV}/\hbox{\AA}$
for fixed $(t_0=2.4\, {\rm eV}, U=6.0\, {\rm eV})$.
The horizontal lines give the experimental results for 3BCMU (green dashed)
and 4BCMU (black dotted).
\label{fig:gapsextrapolated-examples}}
\end{figure}

As an example, in Fig.~\ref{fig:gapsextrapolated-examples}
we show the extrapolated energies for fixed
($t_0=2.4\, {\rm eV}$, $U=6.0\, {\rm eV}$), 
as a function of $V=2.0,2.5,3.0\, {\rm eV}$
for the three parameters $\alpha=3.4,3.5,3.6\, {\rm eV}/\hbox{\AA}$.
As expected for our small parameter window, 
we observe a fairly linear dependence of the gaps
on the parameters $V$ and $\alpha$.
It is seen that not all the gaps can be reproduced perfectly
with a single parameter set. In general, for fixed $(t_0,U)$,
the optical gap
in the triplet sector, $\Delta_{\rm opt}^{\rm t}=E_{\rm T^*}-E_{\rm T}$
requires larger values for $(\alpha, V)$ than the other gaps.
Therefore, we have to compromise to find a good parameter
set. 

To this end, we define $V_i^*$ as the value~$V$ for a given
($t_0,\alpha,U$) which reproduces the experimental
gaps~$i$ ($i=1,2,3,4$
for $E_{\rm gap}$, $\Delta_{\rm opt}^{\rm s}$, $\Delta_{\rm st}$,
$\Delta_{\rm opt}^{\rm t}$) for 3BCMU and 4BCMU from table~\ref{tab:BCMU-energies}.
Then, we calculate the joint 
standard deviation $\sigma(t_0,\alpha,U)$ from
\begin{equation}
[\sigma(t_0,\alpha,U)]^2 = \frac{1}{6} \sum_{i>j}[(V_i^*-V_j^*)/{\rm eV}]^2 \; .
\end{equation}
The optimal parameter set minimizes the spread~$\sigma$.
The results for $\sigma(t_0,\alpha,U)$ 
are shown in table~\ref{tab:3BCMU-optimization}
for 3BCMU and for 4BCMU in table~\ref{tab:4BCMU-optimization}, respectively.

\begin{table}[t]
\begin{center}
\begin{tabular}{c|rrr}
 & \multicolumn{3}{c}{$t_0/{\rm eV}$}\\
$U=5.0\, {\rm eV}$ & 2.0 & 2.2 & 2.4\\
\hline
$\alpha=3.4\, {\rm eV}/\hbox{\AA}$  & \hphantom{1}1.96 & 2.79 & 3.22\\
$\alpha=3.5\, {\rm eV}/\hbox{\AA}$  & {\bf 1.68} & 3.12 & 3.67\\
$\alpha=3.6\, {\rm eV}/\hbox{\AA}$  & 2.66 & 3.63 & 4.11
\end{tabular}
\\[6pt]
\begin{tabular}{c|rrr}
 & \multicolumn{3}{c}{$t_0/{\rm eV}$}\\
$U=5.5\, {\rm eV}$ & 2.0 & 2.2 & 2.4\\
\hline
$\alpha=3.4\, {\rm eV}/\hbox{\AA}$ & \hphantom{1}1.91 & {\bf 1.63} & 2.28\\
$\alpha=3.5\, {\rm eV}/\hbox{\AA}$ & 1.96 & 2.14 & 2.85\\
$\alpha=3.6\, {\rm eV}/\hbox{\AA}$ & 2.53 & 2.63 & 3.37
\end{tabular}
\\[6pt]
\begin{tabular}{c|rrr}
 & \multicolumn{3}{c}{$t_0/{\rm eV}$}\\
$U=6.0\, {\rm eV}$ & 2.0 & 2.2 & 2.4\\
\hline
$\alpha=3.4\, {\rm eV}/\hbox{\AA}$ & 14.07 & 1.95 & {\bf 1.55}\\
$\alpha=3.5\, {\rm eV}/\hbox{\AA}$ & 3.56 & 2.06 & 2.11\\
$\alpha=3.6\, {\rm eV}/\hbox{\AA}$ & 3.59 & 2.91 & 3.67
\end{tabular}
\end{center}
\caption{Spread $\sigma$ as a function of $(t_0,\alpha,U)$ for 3BCMU 
where the experimentally observed gaps
are given by $E_{\rm gap}=2.482\, {\rm eV}$, 
$\Delta_{\rm opt}^{\rm s}=1.896\, {\rm eV}$, $\Delta_{\rm st}=1.0\, {\rm eV}$,
and $\Delta_{\rm opt}^{\rm t}=1.360\, {\rm eV}$.
The three best sets $(t_0,\alpha,U)$ are printed in {\bf bold}.
\label{tab:3BCMU-optimization}}

\begin{center}
\begin{tabular}{c|rrr}
 & \multicolumn{3}{c}{$t_0/{\rm eV}$}\\
$U=5.0\, {\rm eV}$ & 2.0 & 2.2 & 2.4 \\
\hline 
$\alpha=3.4\, {\rm eV}/\hbox{\AA}$ & \hphantom{1}2.42 & 3.19 & 3.66\\
$\alpha=3.5\, {\rm eV}/\hbox{\AA}$ & {\bf 2.30} & 3.65 & 4.16\\
$\alpha=3.6\, {\rm eV}/\hbox{\AA}$ & 3.36 & 4.17 & 4.58
\end{tabular}
\\[6pt]
\begin{tabular}{c|rrr}
 & \multicolumn{3}{c}{$t_0/{\rm eV}$}\\
$U=5.5\, {\rm eV}$ & 2.0 & 2.2 & 2.4 \\
\hline
$\alpha=3.4\, {\rm eV}/\hbox{\AA}$ & \hphantom{1}2.85 & {\bf 2.31} & 2.91\\
$\alpha=3.5\, {\rm eV}/\hbox{\AA}$ & 2.67 & 2.87 & 3.45\\
$\alpha=3.6\, {\rm eV}/\hbox{\AA}$ & 3.37 & 3.34 & 3.96
\end{tabular}
\\[6pt]
\begin{tabular}{c|rrr}
 & \multicolumn{3}{c}{$t_0/{\rm eV}$}\\
$U=6.0\, {\rm eV}$ & 2.0 & 2.2 & 2.4 \\
\hline
$\alpha=3.4\, {\rm eV}/\hbox{\AA}$ & 17.89 & 2.51 & {\bf 2.15}\\
$\alpha=3.5\, {\rm eV}/\hbox{\AA}$ & 3.96 & 2.56 & 2.68\\
$\alpha=3.6\, {\rm eV}/\hbox{\AA}$ & 4.07 & 3.64 & 5.12
\end{tabular}
\end{center}
\caption{Spread $\sigma$ as a function of $(t_0,\alpha,U)$ for 4BCMU 
where the experimentally observed gaps
are given by $E_{\rm gap}=2.378\, {\rm eV}$, 
$\Delta_{\rm opt}^{\rm s}=1.810\, {\rm eV}$, $\Delta_{\rm st}=0.95\, {\rm eV}$,
and $\Delta_{\rm opt}^{\rm t}=1.345\, {\rm eV}$.
The three best sets $(t_0,\alpha,U)$ are printed in {\bf bold}.
\label{tab:4BCMU-optimization}}
\end{table}

The two tables~\ref{tab:3BCMU-optimization} and~\ref{tab:4BCMU-optimization}
indicate that the best set for both 3BCMU and 4BCMU is 
$(t_0=2.4\, {\rm eV},\alpha=3.4\, {\rm eV}/\hbox{\AA},U=6\, {\rm eV})$. 
Then, a look at 
Fig.~\ref{fig:gapsextrapolated-examples} shows that 
$V=3\, {\rm eV}$ is the best value for which we have data available.
Therefore, we shall use the set
$(t_0^*=2.4\, {\rm eV},
\alpha^*=3.4\, {\rm eV}/\hbox{\AA}, 
U^*=6\, {\rm eV},
V^*=3\, {\rm eV})$ as our optimal parameter set.
The trend shows that $\alpha$ might even be a bit smaller,
$\alpha\lesssim 3.4\, {\rm eV}/\hbox{\AA}$, and $t_0$ a bit larger,
$t_0\gtrsim 2.4\, {\rm eV}$, for the optimal case.
We plan to perform a more systematic estimate for the optimal
parameter set in the near future.

For our optimal parameter set $(t_0^*=2.4\, {\rm eV},
\alpha^*=3.4\, {\rm eV}/\hbox{\AA}, 
U^*=6\, {\rm eV},
V^*=3\, {\rm eV})$, the 
corresponding theoretical values for the excitation energies are
$E_{\rm gap}=2.45\, {\rm eV}$, $\Delta_{\rm opt}^{\rm s}=2.00\, {\rm eV}$,
$\Delta_{\rm st}=1.00\, {\rm eV}$, and $\Delta_{\rm opt}^{\rm t}=1.25\, {\rm eV}$.
They are also given in table~\ref{tab:BCMU-energies}.

\subsection{Optically dark in-gap triplet states}

In the triplet sector, we find a series
of optically dark states~Y$_l$
($l=1,2,3,4$) at energies just above the triplet ground state.
As indicated in Fig.~\ref{fig:energylevels}, the
optically dark triplet states~Y$_l$ open decay channels 
for the states X$_1$ and X$_2$.

The structure of the optically dark in-gap triplet states
is also interesting from a theoretical point of view.
In a band picture, the triplet ground state with spin component
$S^z=1$ consists of a hole in the $\downarrow$ valence band
and an electron in the $\uparrow$ conduction
band, both at momentum $k=0$. 
Close in energy are the corresponding excitations at finite 
but small momentum~$k$. The dispersion of these excitations is quadratic
as a function of~$k$.

\begin{figure}[t]
\includegraphics[scale=0.42]{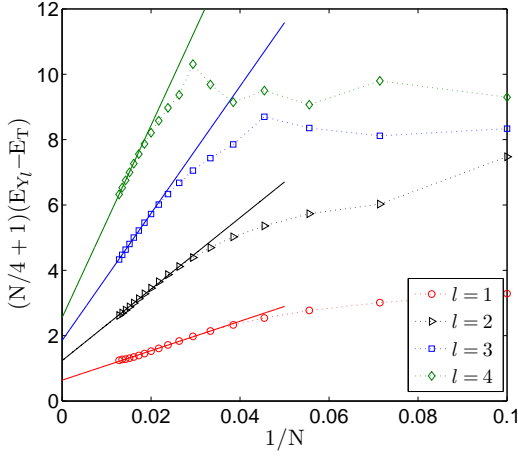}
\caption{(Color online) Scaled excitation energies of
the first four triplet states, $D_l=(N/4+1)(E_{{\rm Y}_l}-E_{\rm T})$, 
as a function of inverse system size $1/N$ for $10\leq N\leq 78$ and
$(t_0^*=2.4\, {\rm eV},
\alpha^*=3.4\, {\rm eV}, 
U^*=6\, {\rm eV},
V^*=3\, {\rm eV})$.
The crossings of the ordinate are equidistant,
$D_1=0.637\, {\rm meV}$, $D_2=1.235\, {\rm meV}$, 
$D_3=1.851\, {\rm meV}$, and $D_4=2.548\, {\rm meV}$, 
 indicating a linear dispersion relation for the excitations.
\label{fig:conformaltower}}
\end{figure}

In the interacting case,
these excitations may form a spin-flip density wave with 
momentum~$q$ whose dispersion relation 
at low-energy is given by
\begin{equation}
\epsilon_{\rm sf}(q) = c_{\rm sf}q \; ,
\end{equation}
where $c_{\rm sf}$ is the (sound) velocity. For finite chains, 
we have quantized quasi-momenta, $q_l= \pi l/[(N/4+1)d]$ ($l=1,2,\ldots$),
where $d$ is the length of the unit cell. 
Therefore, the levels Y$_l$ should obey
\begin{eqnarray}
E_{{\rm Y}_l} - E_{\rm T} &=& \epsilon_{\rm sf}(q_l) + \frac{\alpha_l}{N^2}
+ \ldots \; , \\
\left(\frac{N}{4}+1\right)(E_{{\rm Y}_l} - E_{\rm T}) &=& c_{\rm sf} \frac{l \pi}{d} 
+ \frac{\alpha_l}{N}
+\ldots \; .
\end{eqnarray}
We confirm this hypothesis in Fig.~\ref{fig:conformaltower} where we show
$(N/4+1)(E_{{\rm Y}_l} - E_{\rm T})$ as a function of $1/N$.
As seen from the figure, the energy differences scale
to a finite value linearly in $1/N$. Moreover, 
the extrapolated values are equidistant from which
we can read off $\Delta_{\rm sf}\equiv c_{\rm sf} \pi/d = 0.625\, {\rm meV}$.
Using $d=4.9\, \hbox{\AA}$ we thus estimate the velocity 
for the spin-flip density excitations as $c_{\rm sf}=148\, {\rm m/s}$.
For localized spin models, we have $c_J \approx J (d/4)$ so that,
approximating $c_J\approx c_{\rm sf}$, 
the effective magnetic interaction in our system 
is of the order of one meV, $J\approx (4/\pi)\Delta_{\rm sf}=0.8\, {\rm meV}$.

\subsection{Optically dark in-gap singlet states}

\begin{figure}[tb]
\includegraphics[scale=0.42]{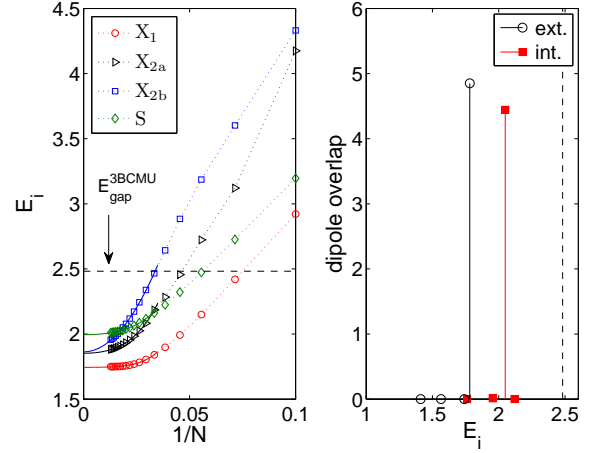}
\caption{(Color online) (a)~Energies of
the lowest singlet states as a function
of inverse system size $1/N$ for $10\leq N\leq 78$ for 
$(t_0^*=2.4\, {\rm eV},
\alpha^*=3.4\, {\rm eV}, 
U^*=6\, {\rm eV},
V^*=3\, {\rm eV})$.
After the extrapolation, the states X$_{2a}$ and X$_{2b}$ are essentially
degenerate, $E_{{\rm X}_1}=1.744\, {\rm eV}$,
$E_{{\rm X}_{2a}}=1.853\, {\rm eV}$,
$E_{{\rm X}_{2b}}=1.863\, {\rm eV}$,
and $E_{{\rm S}}=1.995\, {\rm eV}$.
The horizontal dashed line gives the value for
the charge gap in 3BCMU. 
(b)~Overlap intensity of the dipole operator
for spin-singlet in-gap states after the extrinsic
and the intrinsic relaxation for $N=46$ sites.
\label{fig:darksinglets}}
\end{figure}

In Fig.~\ref{fig:darksinglets}a we show the (relaxed) excitation energies
for the four energetically lowest singlet states as a function 
of $1/N$ for $(t_0^*=2.4\, {\rm eV},
\alpha^*=3.4\, {\rm eV}/\hbox{\AA}, 
U^*=6\, {\rm eV},
V^*=3\, {\rm eV})$.
For our optimal parameter set, 
there are at least three optically dark singlet states below
the singlet exciton. 
We extrapolate $E_{\rm X_1}=1.744\, {\rm eV}$
which is about $0.25\, {\rm eV}$ higher
than estimated from experiment,
see table~\ref{tab:BCMU-energies}.
This indicates that the local correlations could still be larger than 
$U=6\, {\rm eV}$. 
Note, however, that the absolute positions 
of X$_1$ and X$_2$ have not been determined experimentally for nBCMU
but it is known from pump-probe spectroscopy~\cite{Schott-review}
that there are (at least) two optically dark
singlet states below the singlet exciton.

The next two dark in-gap singlets are almost degenerate in energy,
$E_{\rm X_{2a}}=1.853\, {\rm eV}$ and $E_{\rm X_{2b}}=1.863\, {\rm eV}$.
They lie below the singlet exciton, as seen in experiment,
but about $0.15\, {\rm eV}$ higher than estimated in 
table~\ref{tab:BCMU-energies}.
Fig.~\ref{fig:darksinglets}b shows the dipole overlap~\cite{all-of-us}
for the four singlet states X$_1$, X$_{2a}$, X$_{2b}$, and~S
at chain length~$N=46$.
Only the exciton~S has a finite dipole overlap. Note the
energy shift of all in-gap states 
due to the intrinsic Peierls effect.
The Peierls shift amounts to several tenths of an eV.

As in the triplet sector,
we employ the band picture at fixed particle number
to view the elementary excitations 
of the ground state as a hole in the valence band 
and an electron in the conduction band.
In a Wannier picture, the electron-electron interaction
forms bound states from these pairs, such as X$_1$, 
the lowest-lying $A_g$ singlet,
and~S, the singlet exciton with $B_u$ symmetry.
Apart from these bound states, there should be a continuum of
scattering states. The degeneracy of the states X$_{2a}$ and X$_{2b}$
indicates that they are near the threshold to the X-continuum.
Indeed, two-photon absorption above $E_{\rm f}\approx 2.0\, {\rm eV}$
excites states that can fission into two triplets.\cite{Schott-review}
Unfortunately, it takes a significant amount of CPU time
in our DMRG approach to target more than four in-gap singlet states 
simultaneously so that a more detailed
investigation of the in-gap spectrum remains an open problem.

\subsection{Lattice parameters}

\begin{figure}[b]
\includegraphics[scale=0.45]{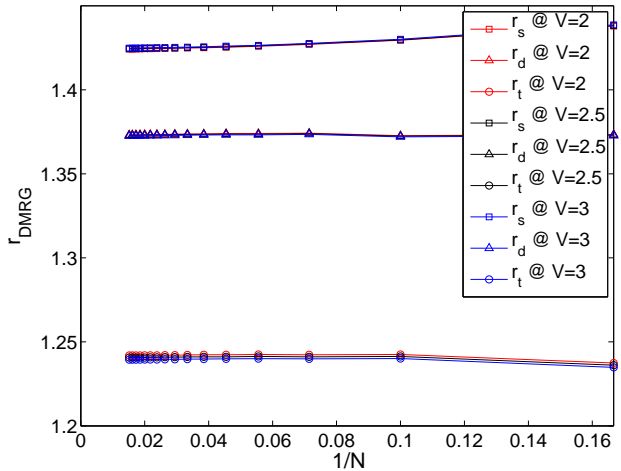}
\caption{(Color online) 
Atomic distances $r_{\rm s}$, $r_{\rm d}$ and $r_{\rm t}$
for the single, double, and triple bonds 
in the chain center as function
of inverse system size $1/N$ for $6\leq N\leq 66$ 
for $(t_0=2.4\, {\rm eV},
\alpha=3.4\, {\rm eV}/\hbox{\AA}, 
U=6\, {\rm eV})$, and $V=2.0,2.5,3.0\, {\rm eV}$.
The extrapolated values for the best parameter set
($V=3\, {\rm eV}$) are $r_{\rm s}=1.425\, \hbox{\AA}$, 
$r_{\rm d}=1.373\, \hbox{\AA}$, and
$r_{\rm t}=1.239\, \hbox{\AA}$. \label{fig:latticeconstants}}
\end{figure}

Finally, we show results for the lattice constants in 
Fig.~\ref{fig:latticeconstants}.
For our optimal set of parameters,
the extrapolated values are $r_{\rm s}=1.425\, \hbox{\AA}$, 
$r_{\rm d}=1.373\, \hbox{\AA}$, and
$r_{\rm t}=1.239\, \hbox{\AA}$,
for the single, double, and triple bond, respectively.
When compared to the experimental values
for PDA single crystals given in Sect.~\ref{subsec:structure},
the values for the single and double bonds are rather good
but the triple bond is too large. 
This can also be seen from the size of the unit cell.
Given the Lewis structure of Fig.~\ref{fig:structure},
the unit cell in chain direction has the length~$d$ with
$d^2=r_{\rm d}^2+(2r_{\rm s}+r_{\rm t})^2-2r_{\rm d}(2r_{\rm s}+r_{\rm t})\cos(\varphi_1)$.
Using $\varphi_1=120^{\circ}$, this results in $d=4.92\, \hbox{\AA}$
which is slightly larger than the experimental values for 3BCMU chains,
$d_{\rm 3BCMU}=4.89\, \hbox{\AA}$.\cite{Schott-review}

The comparison shows that the spring constants~$K$ and~$G$ are too large
and/or the electron-phonon coupling constant $\alpha$ is too small.
Consequently, the spring constants $K$, $G$ should be included
as parameters in the optimization procedure.

\section{Conclusions}
\label{sec:conclusions}

\subsection{Parameter values}

\begin{figure}[htb]
\includegraphics[scale=0.43]{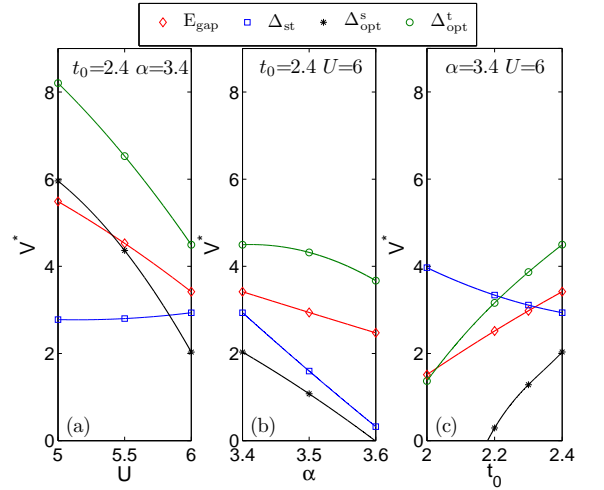}
\caption{(Color online) 
Optimal values $V_i^*$ for the four gaps $E_{\rm gap}$,
$\Delta_{\rm opt}^{\rm s}$, $\Delta_{\rm st}$, and $\Delta_{\rm opt}^{\rm t}$
(a)~as a function of~$U$ for $t_0=2.4\, {\rm eV}$ and
$\alpha=3.4\, {\rm eV}/\hbox{\AA}$,
(b)~as a function of $\alpha$ for $t_0=2.4\, {\rm eV}$ and
$U=6\, {\rm eV}$, and
(c)~as a function of $t_0$ for 
$U=6\, {\rm eV}$ and $\alpha=3.4\, {\rm eV}/\hbox{\AA}$.\label{fig:C1}}
\end{figure}

In our study we use quite sizable values for the Coulomb parameters.
Substantial values for the Hubbard interaction, $U_{\rm CM}=8\, {\rm eV}$,
were used by Chandross, Mazumdar et al.\ in their studies
of the optical properties of poly-phenylene-vinylene 
(PPV) thin films.\cite{Chandross}
For PDA chains immersed in their monomer matrix,
we find $U=6\, {\rm eV}$ and confirm their previous result for the ratio
between $U$ and~$V$, 
$U_{\rm CM}/V_{\rm CM}=\kappa=U/V=2$.\cite{Chandross,Mazumdar-Wang-Zhao}
Note that we additionally screen the long-range part of the Coulomb interaction
by the dielectric constant of the monomer matrix, $\epsilon_d=2.3$.
A further increase of the Hubbard interaction beyond $U=6\, {\rm eV}$
would be problematic, as can be seen from Fig.~\ref{fig:C1}a.
For fixed $t_0$ and $\alpha$, the spread of the gaps
increases with increasing~$U$. In particular, the calculated 
singlet-exciton energy would deviate significantly from
its experimental value.
Moreover, the number of dark singlet states below the exciton would become
larger than expected from experiment.

For fixed $t_0=2.4\, {\rm eV}$ and $U=6\, {\rm eV}$, 
a change from $\alpha=3.4\, {\rm eV}/\hbox{\AA}$ 
would increase the spread in the 
gaps, as can be seen from Fig.~\ref{fig:C1}b.
Therefore, we advocate an electron-phonon coupling strength which
is some 15\% smaller than the value proposed by Ehrenfreund 
et al., $\alpha_{\rm E}=4.0\, {\rm eV}/\hbox{\AA}$.\cite{Ehrenfreund}

In the literature,
typical values for the electron transfer integral
are $t_0=2.5\, {\rm eV}$~\cite{SSH,Girlandoetal}
and $t_0=2.4\, {\rm eV}$~\cite{Chandross,Mazumdar-Wang-Zhao}.  
Our work indicates that a readjustment of the bare electron transfer 
integral is not necessary for nBCMU. 
This can be seen from Fig.~\ref{fig:C1}c, where the spread 
of $V_i$ increases for both smaller and larger~$t_0$.

\subsection{Outlook}

Our study can be extended in several directions.
First, the strength of the spring constants~$K$ and~$G$
must be determined self-consistently.
In this way, a better agreement between theory and experiment
for the lattice parameters can be obtained. 

Second, the parameter search 
in the four-dimensional space $(t_0,\alpha,U,V)$
can be optimized by performing a smooth interpolation
between the 81~data sets which we have investigated numerically thus far.
Moreover, the present analysis does not distinguish
between various ligand types (3BCMU, 4BCMU).
The differences are partly due to strain, and
a distinction between PDAs
must therefore be incorporated in the lattice parameters.
In addition, different side-groups introduce 
(small) electrostatic potential at the carbon atoms which
are linked to the side groups. Our numerical analysis shows that
the influence of an electrostatic potential of the order of 
$\epsilon_i\leq 0.3\, {\rm eV}$ at the side-group sites
changes the spectra by $\delta E\lesssim 0.05\, {\rm eV}$
so that this contribution can be ignored to first approximation.

Third, for an optimal parameter set it is interesting
to study the properties of the ground and excited states in more detail.
For example, the lattice structure of the exciton (polaron-exciton)
and its polarizability can be calculated~\cite{all-of-us,analysis}
and be compared with experiment. Moreover, further in-gap states
should be address, e.g., the spin-2 ground state $1 {}^5A_g$,
or a second singlet exciton $2 {}^1B_{u}^-$.

These tasks are left for a future study.

\subsection*{\bf Acknowledgments}
We thank Gerhard Weiser, Michel Schott, and Benjamin Janesko
for useful discussions.
This work was supported in part
by the Deutsche Forschungsgemeinschaft
through GRK~790, and by the Hungarian Research Fund (OTKA)
Grants Nos.\ K~100908 and K~73455. 
{\"O}.L.\ acknowledges support from the Alexander-von-Humboldt 
foundation and from ETH Zurich
during his time as a visiting professor.

\appendix

\section{Extrinsic dimerization}
\label{appA}

Following Ref.~[\onlinecite{Bursill2003}], we diagonalize
the two-site Peierls--Hubbard-Ohno model in the spin singlet sector
to derive the ground-state energy.
To this end we consider the two states in site representation
\begin{eqnarray}
|1\rangle &=& \sqrt{\frac{1}{2}} 
\left( \hat{c}_{1,\uparrow}^+\hat{c}_{2,\downarrow}^+
-\hat{c}_{1,\downarrow}^+\hat{c}_{2,\uparrow}^+\right) |{\rm vac} \rangle \nonumber \;,
\\
|2\rangle &=& \sqrt{\frac{1}{2}} 
\left( \hat{c}_{1,\uparrow}^+\hat{c}_{1,\downarrow}^+
+\hat{c}_{2,\uparrow}^+\hat{c}_{2,\downarrow}^+\right) |{\rm vac} \rangle \nonumber \;.
\label{appeq:states} 
\end{eqnarray}
In the subspace $S=0$,
the Hamilton matrix of the electronic problem has the entries
$(H_{\rm e})_{i,j}=\langle i | \hat{H}_{\rm e} | j \rangle$ ($i,j=1,2$).
A short calculation gives 
\begin{equation}
\underline{\underline{H_{\rm e}}} = \left(
\begin{array}{cc}
-U/2 & -2T_1(\delta^{\rm e}) \\[3pt]
-2T_1(\delta^{\rm e}) & U/2 -V(\delta^{\rm e})
\end{array}
\label{appeq:matrix}
\right) \; , 
\end{equation}
where
\begin{eqnarray}
V(\delta^{\rm e}) &=&
\frac{V/\epsilon_d}{\sqrt{1+\beta (R_1(\delta^{\rm e})/\mbox{\AA})^2}} \; , 
\nonumber \\
R_1(\delta^{\rm e})&=&r_0-\frac{\delta^{\rm e}}{2\alpha} \; ,\;
T_1(\delta^{\rm e})=t_0+\frac{\delta^{\rm e}}{2} \; .
\end{eqnarray}
$R_1(\delta^{\rm e})$ is the length of the $\sigma$-$p_y$ double bond,
$T_1(\delta^{\rm e})$ is the corresponding electron transfer amplitude,
$U$ is the strength of the electrons' local Coulomb repulsion,
and $V(\delta^{\rm e})$ is their interaction on neighboring sites.
Note that in Ref.~[\onlinecite{Bursill2003}], the unshifted energies
were used, i.e., the previous expressions follow from ours
after an energy shift by $U/2+V(\delta^{\rm e})$.

The ground-state energy $\epsilon_0$ 
of the two-electron system follows from the diagonalization of the
matrix~(\ref{appeq:matrix}) as
\begin{equation}
\epsilon_0(\delta^{\rm e}) = -\frac{1}{2} \left(
V(\delta^{\rm e})
+\sqrt{ [U-V(\delta^{\rm e})]^2+[4 T_1(\delta^{\rm e})]^2 }
\right) \;.
\label{eq:for-gergo}
\end{equation}
Now that we know the ground-state energy of the electronic Hamiltonian 
explicitly, we do not have to invoke the Hellmann--Feynman theorem in order
to determine the optimal values for the electron transfer 
amplitudes~$\delta^{\rm e}$. Instead, we directly minimize
the total ground-state energy
\begin{equation}
E_0(\delta^{\rm e}) = \epsilon_0(\delta^{\rm e}) +
\frac{(\delta^{\rm e})^2 }{4\pi t_0\lambda}
\end{equation}
with respect to $\delta^{\rm e}$ ($\lambda=2\alpha^2/(\pi K t_0)$).
We set
\begin{equation}
V'(\delta^{\rm e})
= \frac{\partial V(\delta^{\rm e})}{\partial \delta^{\rm e}}
= \frac{V\beta R_1(\delta^{\rm e})}{2\alpha \epsilon_d\mbox{\AA}{}^2}
\frac{1}{[1+\beta (R_1(\delta^{\rm e})/\mbox{\AA})^2]^{3/2}} \; .
\end{equation}
Therefore, the optimization of the electron-lattice problem for
two carbon atoms with a double bond leads to
the implicit equation
\begin{equation}
\frac{\delta^{\rm e}}{\pi t_0 \lambda}
= \frac{[2\epsilon_0(\delta^{\rm e})+U]V'(\delta^{\rm e})-8T_1(\delta^{\rm e})}{
2\epsilon_0(\delta^{\rm e})+V(\delta^{\rm e})}\;, 
\label{appeq:implicit}
\end{equation}
which is solved iteratively.

\section{Optical phonons}
\label{app:phonons}

Assuming the Lewis structure of Fig.~\ref{fig:structure},
the position of the atoms in the unit cell are denoted by
the two-dimensional vectors
$\vec{A}_l$, $\vec{B}_l$, $\vec{C}_l$, and $\vec{D}_l$ 
for $l=1,\ldots, L=N/4$.
Their equilibrium positions are denoted as
$\vec{A}_{l,0}$, $\vec{B}_{l,0}$, $\vec{C}_{l,0}$, and $\vec{D}_{l,0}$.
The PDA structure implies
$\vec{B}_{l,0}-\vec{A}_{l,0} =
\vec{D}_{l,0}-\vec{C}_{l,0}=r_{\rm s} \vec{e}_x$,
$\vec{C}_{l,0}-\vec{B}_{l,0}=r_{\rm t} \vec{e}_x$,
and
$\vec{A}_{l+1,0}-\vec{D}_{l,0}= r_{\rm d}(\cos(\phi)\vec{e}_x-\sin(\phi)\vec{e}_y)$ 
for the singlet, triplet, and doublet bonds, where
$\phi=180^{\circ}-\varphi_1=60^{\circ}$.

\subsection{Lagrange function}

We shall treat the atomic motions classically.
For convenience, we use periodic boundary conditions, $L+1\equiv 1$.

The kinetic energy of the atoms is given by
\begin{equation}
T=\frac{M}{2} \sum_{l=1}^L \left[  \left(\dot{\vec{A}_l}\right)^2 
+ \left(\dot{\vec{B}_l}\right)^2 +
\left(\dot{\vec{C}_l}\right)^2 
+\left(\dot{\vec{D}_l}\right)^2 
\right] \; .
\end{equation}
In the spring-constant model, the atoms' potential energy
is approximated by ($g=G/K$)
\begin{eqnarray}
V&=& \frac{K}{2} \sum_{l=1}^L \Biggl[
\left( 
\left| \vec{B}_l -\vec{A}_l \right| -r_{\rm s}
\right)^2
+g
\left( 
\left| \vec{C}_l -\vec{B}_l \right| -r_{\rm t}
\right)^2
\nonumber \\
&& \hphantom{\frac{K}{2}}
+
\left( 
\left| \vec{D}_l -\vec{C}_l \right| -r_{\rm s}
\right)^2
+
\left( 
\left| \vec{A}_{l+1} -\vec{D}_l \right| -r_{\rm d}
\right)^2
\Biggr] \;.\nonumber  \\
\end{eqnarray}
To second order in the displacement $\vec{\delta x}=\vec{x}-\vec{x}_0$
we can write 
\begin{equation}
\left(\left|\vec{x}\right| 
-
\left|\vec{x}{}_0\right|\right)^2
\approx 
\left(  
\vec{\delta x}\cdot \vec{x}_0/|\vec{x}_0|
\right)^2 \; .
\end{equation}
We define the (small) displacements
$\vec{a}_l=\vec{A}_l-\vec{A}_{l,0}= a_l^x\vec{e}_x +a_l^y\vec{e}_y$,
$\vec{b}_l=\vec{B}_l-\vec{B}_{l,0}= b_l^x\vec{e}_x +b_l^y\vec{e}_y$,
$\vec{c}_l=\vec{C}_l-\vec{C}_{l,0}= c_l^x\vec{e}_x +c_l^y\vec{e}_y$, and
$\vec{d}_l=\vec{D}_l-\vec{D}_{l,0}= d_l^x\vec{e}_x +d_l^y\vec{e}_y$.
Then, the Lagrange function 
in the harmonic approximation
becomes $L=T-V$ with
\begin{eqnarray}
T&=&\frac{M}{2} \sum_{l=1}^L \left[  \left(\dot{\vec{a}_l}\right)^2 
+ \left(\dot{\vec{b}_l}\right)^2 +
\left(\dot{\vec{c}_l}\right)^2 
+\left(\dot{\vec{d}_l}\right)^2 
\right] \; , \nonumber \\
V&=& 
\frac{K}{2} \sum_{l=1}^L \biggl[
\left( b_l^x-a_l^x\right)^2
+
g \left( c_l^x-b_l^x\right)^2
+ 
\left( d_l^x-c_l^x\right)^2 \nonumber 
\\
&&
+
\left[ 
\cos(\phi) \left( a_{l+1}^x-d_l^x\right)
- \sin(\phi) \left(a_{l+1}^y-d_l^y\right)
\right]^2 
\biggr] \;. 
\nonumber\\
\end{eqnarray}
The variables $b_l^y$ and $c_l^y$ are cyclic and drop out of the problem.

\subsection{Equations of motion}

The Euler--Lagrange equations can be solved using the Fourier Ansatz
\begin{equation}
x_l(t)=e^{-{\rm i}\omega t}\sum_{k=1}^L\xi_ke^{{\rm i} kl}
\end{equation}
with $k=2\pi m_k/L$, $m_k=0,1,\ldots, L-1$ as the crystal momentum.
Here, $\xi=(\alpha^x,\alpha^y,\beta^x,\gamma^x,
\delta^x,\delta^y)$ corresponds to $x=(a^x,a^y,b^x,c^x,d^x,d^y)$.
The resulting set of six algebraic equations can be cast into a matrix 
equation, ${\cal M}(\omega)\vec{E}_k=(0,0,0,0,0,0)^T$
for the vector $\vec{E}_k=(\alpha_k^x,\alpha_k^y,\beta_k^x,
\gamma_k^x,\delta_k^x,\delta_k^y)^T$. With the abbreviations
$y=M\omega^2/K$, $c=\cos(\phi)$, and $s=\sin(\phi)$, the matrix reads
{\arraycolsep=0pt\begin{eqnarray}
&&{\cal M}(\omega)=\\
&& \!\!\left(
\begin{array}{cccccc}
1+c^2-y & -cs & -1 & 0 & -c^2e^{-{\rm i}k} & c s e^{-{\rm i}k} \\
-cs & s^2-y & 0 & 0 & cs e^{-{\rm i}k} & - s^2 e^{-{\rm i}k} \\
-1 & 0 & 1+g-y & -g & 0 & 0 \\
0 & 0 & -g & 1+g-y & -1 & 0 \\
-c^2 e^{{\rm i}k} & cs e^{{\rm i}k} & 0 & -1 & 1 + c^2-y & -cs  \\
cs e^{{\rm i}k} & -s^2 e^{{\rm i}k} & 0 & 0 & -cs & s^2 -y
\end{array}\!\!
  \right).
\nonumber 
\end{eqnarray}}%
The equation for the vibrational frequencies $\omega_n(k)$ 
as a function of the crystal momentum $k$
results from the characteristic equation,
$\det({\cal M}(\omega))=0$.

We are interested in the optical phonon modes, $\omega_n=\omega_n(k=0)$.
The characteristic equation reduces to
\begin{eqnarray}
y^2(y-2)p(y) &=&0 \; ,\label{thirdorderpolynomial} \\
p(y)&=&y^3-2(g+2)y^2+(7/2+6g)y-3g \nonumber
\end{eqnarray}
for all $g$ and $\phi=60^{\circ}$.
The finite-frequency solutions are denoted as $\omega_n$ ($n=a,b,c,d$).
We set $\omega_b=\sqrt{2K/M}$, and
$\omega_a^2<\omega_c^2<\omega_d^2$ result from the three real roots
of $p(y)=0$ in~(\ref{thirdorderpolynomial}) as 
$\omega_{a,c,d} =\sqrt{Ky_{a,c,d}/M}$.

\section{Bare dispersion relation}
\label{app:barebands}

The operator for the kinetic energy~$\hat{T}$ in eq.~(\ref{eqn:hatT})
is readily diagonalized for periodic boundary conditions,
$N+1\equiv 1$. For the ground state of non-interacting electrons, 
the unit cell consists of four sites, $N=4L$,
and the electron transfer amplitudes follow the periodic pattern
$(t_{\rm s},t_{\rm t},t_{\rm s},t_{\rm d})$. Thus, we may write
\begin{eqnarray}
\hat{T} &=& - \sum_{\sigma}\sum_{n=0}^{L-1} \biggl( 
t_{s} \hat{c}_{4n+1,\sigma}^+\hat{c}_{4n+2,\sigma}
+
t_{t} \hat{c}_{4n+2,\sigma}^+\hat{c}_{4n+3,\sigma}
\nonumber \\
&& 
+
t_{s} \hat{c}_{4n+3,\sigma}^+\hat{c}_{4n+4,\sigma}
+ 
t_{d} \hat{c}_{4n+4,\sigma}^+\hat{c}_{4(n+1)+1,\sigma}\biggr) 
+ {\rm h.c.} \; .\nonumber \\
\label{eqn:Tperiodic}
\end{eqnarray}
We introduce the four operators $\hat{b}_{M;k,\sigma}$ 
for electrons with quasi-momentum 
$k=-\pi+2\pi m_k/L$, $m_k=0,1,\ldots,L-1$ via
\begin{equation}
\hat{b}_{M;k,\sigma} = \sqrt{\frac{1}{L}} \sum_{n=0}^{L-1} e^{-\I k n}
\hat{c}_{4n+M,\sigma} \; .
\end{equation}
The inverse transformation reads
\begin{equation}
\hat{c}_{4n+M,\sigma} 
= \sqrt{\frac{1}{L}} \sum_{k} e^{\I k n} \hat{b}_{M;k,\sigma} 
\; , \label{trafo}
\end{equation}
where we used that the sum over the quasi-momenta~$k$
generates the orthogonality relation 
\begin{equation}
\frac{1}{L}\sum_{k} e^{\I k(m-n)} = \delta_{m,n} 
\end{equation}
for two lattice indices $m,n$. In turn,
the sum over lattice indices~$n$ 
leads to the orthogonality relation
\begin{equation}
\frac{1}{L}\sum_{n=0}^{L-1}e^{\I n(k-p)} = \delta_{k,p}
\end{equation}
for two quasi-momenta $k,p$.

When we apply the transformation~(\ref{trafo}) to the kinetic 
energy~(\ref{eqn:Tperiodic}), we obtain
\begin{eqnarray}
\hat{T}&=& -\sum_{k,\sigma} \biggl( 
t_{\rm s} \hat{b}_{1;k,\sigma}^+\hat{b}_{2;k,\sigma}^{\vphantom{+}}
+
t_{\rm t} \hat{b}_{2;k,\sigma}^+\hat{b}_{3;k,\sigma}^{\vphantom{+}}
\nonumber \\
&& +
t_{\rm s} \hat{b}_{3;k,\sigma}^+\hat{b}_{4;k,\sigma}^{\vphantom{+}}
+
t_{\rm d} e^{\I k}\hat{b}_{4;k,\sigma}^+\hat{b}_{1;k,\sigma}^{\vphantom{+}}\biggr)
+ {\rm h.c.} \; .
\end{eqnarray}
The remaining task is to band-diagonalize the kinetic energy.
To this end we diagonalize the $4\times 4$-matrix ${\cal M}_k$ with 
\begin{equation}
{\cal M}_k = \left(
\begin{array}{cccc}
0 & -t_{\rm s} & 0 & -t_{\rm d} e^{-\I k} \\
-t_{\rm s} & 0 & -t_{\rm t} & 0 \\
0 & -t_{\rm t} & 0 & -t_{\rm s} \\
-t_{\rm d} e^{\I k} & 0 & -t_{\rm s} & 0
\end{array}
  \right) \; .
\end{equation}
Its eigenvalues in ascending order are $E_1(k)=-\epsilon_2(k)$,
$E_2(k)=-\epsilon_1(k)$,
$E_3(k)=\epsilon_1(k)$,
$E_4(k)=\epsilon_2(k)$ with $\epsilon_1(k)<\epsilon_2(k)$.
We find~\cite{Bursill2001}
{\arraycolsep=2pt\begin{eqnarray}
[\epsilon_{1,2}(k)]^2 &=& 
t_{\rm s}^2+t_{\rm d}^2/2+t_{\rm t}^2/2
\nonumber \\
&& \pm 
\sqrt{(t_{\rm d}^2-t_{\rm t}^2)^2/4+t_{\rm s}^2
\left[t_{\rm d}^2+t_{\rm t}^2+2t_{\rm d}t_{\rm t}\cos(k)\right]}
\nonumber \; . \\
\label{appeq:eps}
\end{eqnarray}}%
We write
\begin{equation}
{\cal M}_k = {\cal U}_k^+ {\cal D}_k {\cal U}_k\; ,
\end{equation}
where ${\cal D}={\rm diag}(-\epsilon_2(k),-\epsilon_1(k),
\epsilon_1(k),\epsilon_2(k))$ is a diagonal matrix which contains the
eigenvalues of ${\cal M}_k$.
We define the band-diagonal operators
\begin{equation}
\left(
\begin{array}{c}
\hat{\alpha}_{k,\sigma} \\
\hat{\beta}_{k,\sigma} \\
\hat{\gamma}_{k,\sigma} \\
\hat{\delta}_{k,\sigma} 
\end{array}
\right)
= {\cal U}_k 
\left(
\begin{array}{c}
\hat{b}_{1;k,\sigma} \\
\hat{b}_{2;k,\sigma} \\
\hat{b}_{3;k,\sigma} \\
\hat{b}_{4;k,\sigma} 
\end{array}
\right) 
\end{equation}
and find
\begin{eqnarray}
\hat{T}&=& \sum_{k,\sigma} \biggl[ 
\epsilon_2(k)
\left(
\hat{\delta}_{k,\sigma}^+\hat{\delta}_{k,\sigma}^{\vphantom{+}}
-
\hat{\alpha}_{k,\sigma}^+\hat{\alpha}_{k,\sigma}^{\vphantom{+}}
\right)\nonumber \\
&& \hphantom{\sum_{k,\sigma} \biggl[ }
+
\epsilon_1(k)
\left(
\hat{\gamma}_{k,\sigma}^+\hat{\gamma}_{k,\sigma}^{\vphantom{+}}
-
\hat{\beta}_{k,\sigma}^+\hat{\beta}_{k,\sigma}^{\vphantom{+}}
\right)\biggr]
\; .
\end{eqnarray}
The $\alpha$- and $\beta$-bands are the valence bands which are
filled in the ground state at half band-filling.
The $\gamma$- and $\delta$-bands are the conduction bands which are
empty in the half-filled ground state.

The bare gap at half band-filling obeys $\Delta^{\rm bare} = 2 \epsilon_1(0)$
with
\begin{equation}
[\epsilon_1(0)]^2 =  
t_{\rm s}^2+t_{\rm d}^2/2+t_{\rm t}^2/2 -
\sqrt{4t_{\rm s}^2+(t_{\rm t}-t_{\rm d})^2}(t_{\rm d}+t_{\rm t})/2
\; .
\end{equation}
For small deviations, $t_{\rm s},t_{\rm d},t_{\rm s}\approx t_0$, 
this simplifies to~\cite{Bursill2001}
\begin{equation}
\Delta^{\rm bare} \approx \left| t_{\rm t}+t_{\rm d}-2t_{\rm s}\right|\; .
\end{equation}


\begin{thebibliography}{99}

\bibitem{PDAreview} {\sl Polydiacetylenes}, ed.~by H.-J.\ Cantow
(Advances in Polymer Sciences~{\bf 63}, Springer, Heidelberg, 1984);
{\sl Polydiacetylenes}, ed.\ by D.\ Bloor and R.R.\ Chance
(Nijhoff, Dordrecht, 1985).

\bibitem{Sariciftci} {\sl Primary photoexcitations
in conjugated polymers}, ed.\ by N.S.\ Sariciftci
(World Scientific, Singapore, 1997).

\bibitem{Schott-review} M.\ Schott, in 
{\sl Photophysics of molecular materials: from single molecules 
to single crystals}, ed.\ by G.\ Lanzani
(Wiley-VCH, Weinheim, 2006), p.~49.

\bibitem{Schottzucht} S.\ Spagnoli, J.\ Berrkhar, C.\ Lapersonne-Meyer, 
and M.\ Schott, J.~Chem.\ Phys.~{\bf 100}, 6195 (1994).

\bibitem{Schottexciton} F.\ Dubin, R.\ Melet, T.\ Barisien, R.\ Grousson, 
L.\ Legrand, M.\ Schott, and V.\ Voliotist,
Nature Physics~{\bf 2}, 32 (2006).

\bibitem{Parry} D.E.~Parry, Chem.\ Phys.\ Lett.~{\bf 43}, 597 (1976);
Chem.\ Phys.\ Lett.~{\bf 46}, 605 (1977).

\bibitem{YangandKertesz} S.~Yang and M.\ Kertesz, 
J.\ Phys.\ Chem.\ A~{\bf 110}, 9771 (2006).

\bibitem{Janesko-paper} B.G.~Janesko, 
J.~Chem.~Phys.~{\bf 134}, 184105 (2011).

\bibitem{Janesko-private} B.G.~Janesko, private communication (2011).

\bibitem{LouieRohlfing} M.~Rohlfing and S.G.~Louie,
Phys.~Rev.~Lett.~{\bf 82}, 1959 (1999).

\bibitem{LDA-GW-BSE} J.-W.\ van der Horst, P.A.~Bobbert,
M.A.J.\ Michels, G.\ Brocks, and P.J.\ Kelly,
Phys.\ Rev.\ Lett.~{\bf 83}, 4413 (1999);
J.-W.\ van der Horst, P.A.\ Bobbert, P.H.L.\ de Jong,
M.A.J.\ Michels, G.\ Brocks, and P.J.\ Kelly,
Phys.\ Rev.\ B~{\bf 61}, 15817 (2000);
J.-W.\ van der Horst, P.A.\ Bobbert, and M.A.J.\ Michels,
Phys.\ Rev.\ B~{\bf 66}, 035206 (2002).

\bibitem{van-der-Horst} J.-W.\ van der Horst, P.A.~Bobbert,
M.A.J.~Michels, and H.~B\"a\ss ler, J.\ Chem.\ Phys.~{\bf 114},
6950 (2001).

\bibitem{Rissler-Grage} A.\ Grage, F.\ Gebhard, and J.\ Rissler, 
J.\ Stat.\ Mech.\ Exp.\ Theor.~P08009 (2005).

\bibitem{PPP} R.~Pariser and R.G.~Parr,
J.~Chem.\ Phys.~{\bf 21}, 466 (1953);
J.A.~Pople, Trans.\ Farad.\ Soc.~{\bf 49}, 1375 (1953).

\bibitem{steve} S.R.~White, Phys.~Rev.~Lett.~{\bf 69}, 2863
(1992); Phys.\ Rev.\ B~{\bf 48}, 10345 (1993).

\bibitem{all-of-us} G.\ Barcza,  \"O.\ Legeza, F.\ Gebhard, and 
R.M.\ Noack, Phys.\ Rev.\ B~{\bf 81}, 045103 (2010).

\bibitem{Bursill2001} A.~Race, W.~Barford, and R.J.~Bursill,
Phys.~Rev.~B~{\bf 64}, 035208 (2001).

\bibitem{Bursill2003} A.~Race, W.~Barford, and R.J.~Bursill,
Phys.\ Rev.\ B~{\bf 67}, 245202 (2003).

\bibitem{Ohno} K.~Ohno, Theor.\ Chim.\ Acta~{\bf 2}, 219 (1964).

\bibitem{Chandross} M.~Chandross, S.~Mazumdar, M.~Liess, P.A.~Lane,
Z.V.\ Vardeny, M.\ Hamaguchi, and K.\ Yoshino,
Phys.\ Rev.\ B~{\bf 55}, 1486 (1997);
M.~Chandross and S.~Mazumdar, Phys.\ Rev.\ B~{\bf 55}, 1497 (1997).

\bibitem{KobeltPaulus} T.D.~Kobelt and E.F.~Paulus, 
Acta Crystallogr. {\bf B~30}, 232 (1974).

\bibitem{ChemlaZyss} For a review, see
M.\ Schott and G.\ Wegener, in {\sl Nonlinear Optical 
Properties of Organic Molecules and Crystals}, Vol.~2, ed.\ by
D.S.\ Chemla and J.\ Zyss (Academic Press Inc., London, 1987), p.~3.

\bibitem{SpagnoliRaman} S.~Spagnoli, J.~Berrehar, J.-L.\ Fave, and
M.~Schott, Chem.\ Phys.~{\bf 333}, 254 (2007).

\bibitem{HorvathWeiserLapersonne} A.\ Horvath, G.\ Weiser, 
C.\ Lapersonne-Meyer, M.\ Schott, and S.\ Spagnoli,
Phys.\ Rev.~B~{\bf 53}, 13507 (1996).

\bibitem{Weiserprivate} G.\ Weiser, private communication (2012).

\bibitem{Loudon} R.~Loudon, Am.\ J.\ Phys.~{\bf 27}, 649 (1959).

\bibitem{KochundCo} L.~B\'anyai, I.~Galbraith, C.~Ell, and H.~Haug,
Phys.\ Rev.\ B~{\bf 36}, 6099 (1987).

\bibitem{SSH} A.J.\ Heeger, S.\ Kivelson, J.R.\ Schrieffer, and W.-P.\ Su,
Rev.\ Mod.\ Phys.~{\bf 60}, 781 (1988).

\bibitem{Ehrenfreund} E.\ Ehrenfreund, Z.\ Vardeny, O.\ Brafman, 
and B.\ Horovitz,
Phys.\ Rev.~B~{\bf 36}, 1535 (1987).

\bibitem{Solyom3} See, for example, J.~S\'olyom,
{\sl Fundamentals of the Physics of Solids}, Vol.~3
(Springer, Berlin, 2010), chap.~30.

\bibitem{Weiser} G.~Weiser, Phys.~Rev.~B~{\bf 45}, 14076 (1992).

\bibitem{dbss-1} \"O.\ Legeza, J.\ R\"oder, and B.A.\ Hess,
Phys.\ Rev.\ B~{\bf 67}, 125114 (2003).

\bibitem{dbss-2} \"O.\ Legeza and J.\ S\'olyom,
Phys.\ Rev.\ B~{\bf 70},  205118  (2004).
 
\bibitem{barcza} G.\ Barcza, \"O.\ Legeza, K.H.\ Marti, and M.\ Reiher, 
Phys.\ Rev.\ A~{\bf 83}, 012508 (2011). 

\bibitem{SutherlandDennison} G.B.B.M.\ Sutherland and D.M.\
Dennison, Proc.\ Roy.\ Soc.\ A~{\bf 148}, 250 (1935).

\bibitem{VallanceJones} A.V.\ Jones, 
Proc.\ Roy.\ Soc.~A~{\bf 211}, 285 (1952).

\bibitem{Girlandoetal} A.\ Girlando, A.\ Painelli, G.W.\ Hayden, 
and Z.G.\ Soos, Chem.\ Phys.~{\bf 184}, 139 (1994).

\bibitem{Herzberg} G.\ Herzberg, {\sl Infrared and Raman Spectroscopy of
polyatomic molecules\/} (van Nostrand, New York, 1945).

\bibitem{PAdistortion} C.S.~Yannoni and T.C.~Clarke,
Phys.\ Rev.\ Lett.~{\bf 51}, 1191 (1983).

\bibitem{Barcza-unpublished} G.\ Barcza, PhD thesis (Budapest, 2013,
unpublished).

\bibitem{Mazumdar-Wang-Zhao} S.~Mazumdar, Z.~Wang, and H.~Zhao,
in {\sl Ultrafast Dynamics and Laser Action of Organic Semiconductors},
ed.~by Z.V.\ Vardeny (CRC Press, Boca Raton, USA, 2009), p.~77.

\bibitem{analysis} J.~Rissler, H.~B\"{a}\ss ler, F.~Gebhard, and
P.~Schwerdtfeger, Phys.~Rev.~B~{\bf 64}, 045122 (2001); J.\ Rissler,
F.~Gebhard, and E.~Jeckelmann, J.~Phys.~Cond.~Matt.~{\bf 17}, 4093 (2003).


\end{thebibliography}
\end{document}